\documentclass[traditabstract]{aa} % for the abstract without structuration 

\usepackage{graphicx}
\usepackage{txfonts}
\usepackage{lipsum}
\usepackage{subcaption}         % necessary for continued figures, example in section 3
\usepackage{lscape}             % to rotate a single page table, example in appendix.
\usepackage{placeins}           % useful with \FloatBarrier, to keep 
\usepackage{graphicx,dblfloatfix}
\usepackage{gensymb}
\usepackage{booktabs}
\usepackage{amsmath}
\usepackage{orcidlink}
\usepackage{natbib}
\usepackage{float}
\usepackage{mwe}
\usepackage{txfonts}
\usepackage{hyperref}
\hypersetup{
     colorlinks   = true,
     citecolor    = blue
}

\usepackage{placeins}
\usepackage{siunitx}
\usepackage{gensymb}
\usepackage{caption,subcaption}
\usepackage{subcaption}
\usepackage{pgffor}
\usepackage{dblfloatfix}    % To enable figures at the bottom of page

\usepackage{lscape} % for 'landscape' environment
\usepackage{multirow}
\usepackage[export]{adjustbox}
\usepackage{numprint}
\npthousandsep{\,}

\usepackage[normalem]{ulem}

\usepackage{tabularx}

\newcommand\clearrow{\global\let\rowmac\relax}
\clearrow

\def \msun{$\,M_{\odot}\,$}

\def \magperarcsec{mag arcsec$^{-2}$}
\def \Hi{\ion{H}{i}}

\def \msunpcsq{M$_{\odot}$\,pc$^{-2}$}

\makeatletter
\AtBeginDocument{%
  \nolinenumbers
  \let\switchlinenumbers\relax
  
  \let\runningpagewiselinenumbers\relax
  \let\modulolinenumbers\relax
  \let\linenumbers\relax
  \let\resetlinenumber\relax
  \let\setrunninglinenumbers\relax
  \let\setpagewiselinenumbers\relax
  
  \let\makeLineNumber\relax
  \let\LineNumber\relax
  \ifx\linenomath\undefined\else
    \let\linenomath\relax
    \let\endlinenomath\relax
    \let\linenomathWithnumbers\relax
    
  \fi
}
\makeatother

\begin{document}

    \title{Deep imaging of the galaxy Malin~2 shows new faint structures and a candidate satellite dwarf galaxy}
    
    \titlerunning{Deep imaging of the galaxy Malin~2 shows new faint structures}

   \author{Junais\inst{1,2}, Ignacio Ruiz Cejudo\inst{1,2}, Sergio Guerra Arencibia\inst{1,2}, Ignacio Trujillo\inst{1,2}, Miguel R. Alarcon\inst{1,2}, Miquel Serra-Ricart\inst{1,2,3}, Johan H. Knapen\inst{1,2}, Pierre-Alain Duc\inst{4}}
    \authorrunning{Junais et al.}

    \institute{
    Instituto de Astrof\'{i}sica de Canarias, V\'{i}a L\'{a}ctea S/N, E-38205 La Laguna, Spain 
    \and
    Departamento de Astrof\'{i}sica, Universidad de La Laguna, E-38206 La Laguna, Spain
    \and
    Light Bridges S.L., Observatorio del Teide, Carretera del Observatorio s/n, E-38500 Guimar, Tenerife, Canarias, Spain
    \and
    Université de Strasbourg, CNRS, Observatoire astronomique de Strasbourg, UMR 7550, 67000 Strasbourg, France
    }
   \date{Received xx, 2025; accepted xx, 2025}

% \abstract{}{}{}{}{}
% 5 {} token are mandatory

\abstract
{Giant Low Surface Brightness (GLSB) galaxies represent an extreme class of disk galaxies characterized by exceptionally large sizes and low stellar densities. Their formation and evolutionary pathways remain poorly constrained, primarily due to the observational challenges associated with detecting their faint stellar disks. In this work, we present deep, multi-band optical imaging of Malin~2, a prototypical GLSB galaxy, obtained with the newly commissioned Two-meter Twin Telescope (TTT) at the Teide Observatory. Our \textit{g}-, \textit{r}-, and \textit{i}-band observations reach surface brightness depths of 30.3, 29.5, and 28.2\,\magperarcsec{} (3$\sigma$, in areas equivalent to $10^{\prime\prime} \times 10^{\prime\prime}$), respectively, enabling us to trace the stellar disk of Malin~2 out to a radius of $\sim$110\,kpc for the first time. We observe new diffuse stellar structures, including a prominent stellar emission toward the northwest region of Malin~2. This emission coincides well with the \Hi{} gas distribution in this region. We also identify a faint spiral arm-like structure in the southeast of Malin~2. Additionally, we report the discovery of a very faint dwarf galaxy, TTT-d1 ($\mu_{0,g} \sim 26$\,\magperarcsec{}), located at a projected distance of $\sim$130\,kpc southeast of Malin~2. If physically associated with Malin~2, it would represent the first known satellite ultra-diffuse galaxy of a GLSB galaxy. We perform a multi-directional wedge photometric analysis of Malin~2 and find that the galaxy has significant azimuthal variations in its stellar emission. A comparison of the stellar mass surface density profiles of Malin~2 with those of a large sample of nearby spiral galaxies and other GLSB galaxies shows that Malin~2 lies at the extreme end of both these classes of objects in its radial extent and stellar mass surface density distribution. The spatial overlap between the asymmetric stellar emission and a lopsided \Hi{} distribution suggests that Malin~2’s giant LSB disk has contributions from tidal interactions. Our results highlight the importance of ultra-deep, wide-field imaging in understanding the structural complexity of GLSB galaxies. Upcoming surveys such as LSST will be crucial to determine whether the features we observe in Malin~2 are common to other giant LSB disk galaxies.}


   \maketitle

%%%%%%%%%%%%%%%%%%%%%%%%%%%%%%%%%%%%%%%%%%%%%%%%%%%%%%%%%%%%%%
\section{Introduction}

Low-surface-brightness (LSB) galaxies are diffuse galaxies that are much fainter than the typical night sky surface brightness.
The extreme faintness of LSB galaxies hinders in-depth observations even in the local universe, and a large fraction of such galaxies is missing from most observations to date \citep{Martin2019}. In recent years, with the advent of powerful instruments, it has become possible to obtain deeper observations, allowing astronomers to study these galaxies with a new perspective.

LSB galaxies span a wide range of galaxy sizes, masses, and morphologies, from extreme size down to the more common dwarf galaxies. Giant LSB (GLSB) galaxies are an extreme case of LSB galaxies, with a huge LSB disk (scale length of 10 to 50\,kpc; \citealt{Sprayberry1995}) and rich in gas content (M$_{\mathrm{HI}} > 10^{10}$ \msun{}; \citealt{Matthews2001}). \citet{Sprayberry1995} classifies GLSB disk galaxies using a diffuseness index criterion based on the \textit{B}-band disk central surface brightness ($\mu_{0,B}$) and scale length ($R_{\mathrm{s}}$). According to this definition, GLSB disk galaxies have a high diffuseness index, following the relation $\mu_{0,B} + 5\log(R_{\mathrm{s}}) > 27$, where $\mu_{0,B}$ is \magperarcsec{} and $R_{\mathrm{s}}$ in kpc. The prototype of the class of GLSB galaxies is Malin~1, discovered by \citet{Bothun1987}. The inner region of Malin~1 is an early-type high surface brightness spiral (SB0/a), surrounded by a huge low surface brightness disk extending at least $\sim120$\,kpc in radius, making it the largest spiral galaxy known \citep{moore2006,Barth2007}. The origin of such GLSB galaxies is still debated, with several formation scenarios proposed (e.g., major mergers, high-spin dark matter (DM) haloes, low-density environments, satellite accretion, and head-on collisions; \citealt{Saburova2021}). GLSB galaxies are thought to be rare due to their unusual disk extent and faintness. However, a recent work by \citet{Saburova2023} predicted that there could be about 13000 GLSB galaxies just within $z<0.1$. Until now, only 107 GLSB galaxies have ever been discovered \citep{Zhu2023}, indicating that a full census of these giants remains incomplete. 

In this paper, we focus on Malin~2 (also known as F568-6), another prototypical GLSB galaxy like Malin~1. The morphological type of Malin~2 is that of a late-type (Scd) spiral with a huge extended disk \citep{Kasparova2014}. \citet{Sprayberry1995} classifies Malin~2 as a giant LSB disk with a \textit{B}-band central disk surface brightness ($\mu_{0,B}$) of 23.4\,\magperarcsec{} and a disk scale length ($R_{\mathrm{s}}$) of 22.6\,kpc. Malin~2 has an huge \Hi{} disk extending out to a radius of $\sim120$\,kpc, rich in gas content with $M_{\rm HI}\sim6\times10^{10},M_{\odot}$, and a molecular gas fraction of up to 2\% percent of the \Hi{} mass \citep{Pickering1997,Das2010}. Using long-slit spectroscopic observations and mass modeling, \citet{Kasparova2014} suggested that the giant disk of Malin~2 was likely formed due to a sparse and shallow DM halo, rather than any catastrophic scenarios like major mergers. However, to rule out such scenarios, we also need deep imaging observations that can reveal the LSB signature of any past interactions, if any. In the case of Malin~1, deep imaging observations by \citet{Galaz2015} showed very low surface brightness tidal features linking Malin~1 and a neighboring galaxy, indicating a likely past interaction between the two. However, in the case of Malin~2, it has never been imaged to surface brightness depths sufficient to trace such features. Existing optical data for Malin~2 such as the Sloan Digital
Sky Survey (SDSS) and the Dark Energy Camera Legacy Survey (DECaLS) reach only down to a surface brightness depth of $\sim$28\,\magperarcsec{} in the optical bands, leaving the outer regions of the stellar disk of Malin~2 largely unexplored.

In this work, we present the deepest optical imaging of Malin~2 to date, probing $\sim$2\,mag arcsec$^{-2}$ deeper than previous studies. We explore the previously unseen low surface brightness stellar emission from the giant LSB disk of Malin~2 to shed light on its true size and formation. Throughout this paper we adopt a flat $\Lambda$CDM cosmology with $H_0=70$ km s$^{-1}$ Mpc$^{-1}$, $\Omega_M=0.27$, and $\Omega_\Lambda=0.73$. This corresponds to a projected angular scale of 0.904\,kpc\,arcsec$^{-1}$ for Malin~2 at a luminosity distance of 204.1 Mpc ($z = 0.046$).

The paper is structured as follows: Sect. \ref{sect:data} provides an overview of the data, observations, and reduction process. Section \ref{sect:results} focuses on the analysis and results. Sections \ref{sect:discussion} and \ref{sect:conclusion} present the discussion and conclusions, respectively.

%%%%%%%%%%%%%%%%%%%%%%%%%%%%%%%%%%%%%%%%%%%%%%%%%%%%%%%%%%%%%%
\section{Data} \label{sect:data}

We obtain the deep imaging of Malin~2 presented in this work using one of the newly commissioned Two-meter Twin Telescope (TTT), a Ritchey-Chrétien 2-m telescope located at Teide Observatory, Canary islands (Lat. $28^\circ 18^{\prime} 04^{\prime\prime}$ N, Long. $16^\circ 30^{\prime} 38^{\prime\prime}$W). The camera used for the observations is a QHY600M Pro camera (sCMOS BSI detector) covering a field of view of $9.6^{\prime}$ x $6.5^{\prime}$ and a pixel scale of 0.06$^{\prime\prime}$ $\rm{pix}^{-1}$, although the data is re-binned in 3x3 pixels, thus resulting in a pixel scale of 0.194$^{\prime\prime}$ $\rm{pix}^{-1}$. As this is the first paper using the TTT data, in the following subsections, we provide a detailed description of the observations and the data reduction.

We observe Malin~2 between 25-03-2025 and 31-03-2025, using $g^\prime$, $r^\prime$, and $i^\prime$-band standard Sloan filters manufactured by Baader Planetarium (hereupon, we refer to them simply as \textit{g}, \textit{r} and \textit{i-}band filters). The total exposure times are 7.2 hours (\textit{g}-band), 7.0 hours (\textit{r}-band), and 3.2 hours (\textit{i}-band), with a typical measured full width half-maximum (FWHM) in the images of 1.8\arcsec{} for point-like sources and a typical sky brightness of 21.75, 20.75, and 20.35\,\magperarcsec{} for \textit{g}-, \textit{r}-, and \textit{i}-band filters, respectively. 

The limiting $3\sigma$ surface brightness levels of our observations measured in regions equivalent to an area of $10^{\prime\prime}\times10^{\prime\prime}$ following Appendix A of \citet{Roman2020} are 30.3 (\textit{g}-band), 29.5 (\textit{r}-band), and 28.2 (\textit{i}-band) \magperarcsec{}. Comparing these depths with the deep imaging surveys like DECaLS DR10, we reach 1.5 magnitudes deeper in the \textit{g}- and \textit{r}-bands and 0.5 magnitude deeper in \textit{i}-band in the Malin~2 field (DECaLS DR10, $(3\sigma, 10^{\prime\prime}\times10^{\prime\prime}) \sim$ 28.8, 28.1 and 27.7\,\magperarcsec{}).

\subsection{Observational strategy}\label{sec:obstrategy}

The strategy we use for performing the observations aims to maximize the time spent observing Malin~2 and, at the same time, accurately characterize the illumination of the sky at the moment of the data collection. The sky characterization is fundamental in order to preserve the LSB structure and not to introduce artificial structure in the images. Part of this sky characterization is done in the flat-field correction, and as described in Sect. \ref{sect:bias_flat_field_correction}, we perform this correction using the scientific data itself, which allows us to correct the inhomogeneous sensitivity of the detector (as a typical flat-field correction does), but without introducing the gradients present in the twilight/dome flats and correcting the illumination at the moment of the observations. However, in order to build this flat-field with the scientific exposures, we need a large-displacement dithering pattern \citep{Trujillo2016}. Thus, we build a dithering pattern with offsets of up to 6\arcmin{}, which is larger than the size of Malin~2 ($\sim4$\arcmin{}), ensuring that there are exposures in which every pixel of the detector is free of sources and that we are always integrating on the galaxy. Additionally, due to the virtually non-existent read-out time (CMOS detector), we are able to efficiently do exposures of 1 minute, which contribute to minimizing saturation and get a precise characterization of the sky.

\subsection{Data reduction}
We reduce the data using a pipeline specifically developed for preserving the LSB features of the data and minimizing the contamination present in the region of the observation (mainly gradients produced by the sky brightness). We describe the main steps of the data reduction process in the following subsections.

\subsubsection{Bias, Dark and Flat-field correction}\label{sect:bias_flat_field_correction}
The first step in the reduction of the data is to cope with saturated pixels and perform the bias correction. We solve the first of these tasks by masking all the pixels above 65500 Analog Digital Units (ADU), which in the data are located at the central regions of some bright sources. We then apply the bias correction by subtracting a master bias (resulting from combining a set of individual biases) from the frames. Dark current is negligible in the detector for the exposure times we use, so we do not apply a correction to avoid introducing additional noise \citep{Alarcon2023}.

As outlined in Sec ~\ref{sec:obstrategy}, we design the observation strategy to allow performing the flat-field correction using science images, following a similar strategy as that described in \citet{Trujillo2021} and \citet{Zaritsky2024}. First, we normalize science images (bias-corrected) using the resistant mean of the values inside a ring that defines a constant-illumination section, centered on the center of the CCD, with an inner radius of 650 pixels and a width of 200 pixels. Then, we combine the normalised, bias-subtracted science images to build a flat. However, instead of using all the images to build a master flat, for each image, we use the ones that are close in time. This strategy allows us to better characterize the sky illumination at the moment of the observation. The size of the window of frames used for this \textit{Running Flat} strategy is to be 30 (i.e., we select the 15 previous and following images). \\

We iterate this strategy for getting the flat-field three times to improve the quality of the results. On the first one, the flat is done by combining science images with a sigma clipping median stacking. On the second and third iterations, we first mask all the detected sources using \texttt{Gnuastro}\footnote{\url{https://www.gnu.org/software/gnuastro}} \texttt{NoiseChisel} \citep{Akhlaghi2015,Akhlaghi2019}, and then we median-combine the normalised, masked images. Each of these iterations produces a better flat field, allowing us to improve the quality of the data and build better masks, thus further improving the flat itself. Finally, we divide the individual science images by their corresponding flat. Additionally, to deal with vignetting problems towards the corner of the detectors, we remove all the pixels where the masterflats have an illumination lower than 60$\%$ of the central pixels.

\subsubsection{Astrometry}

To determine the astrometry of the individual science images, we calculate a first astrometric solution using \texttt{Astrometry.net} \citep[v0.94;][]{Lang2010}, with \textit{Gaia} DR3 \citep{Gaia2021} as our reference catalog. We then improve this first astrometric solution using \texttt{SCAMP} \citep[v.2.10.0;][]{Bertin2006}. \texttt{SCAMP} reads the catalogs generated by \texttt{SExtractor} \citep[v2.25.0;][]{Bertin1996} to compute the distortion of the CCD. We repeat this process three times, as each iteration improves the previous solution. After that, we run \texttt{SWarp} \citep[v2.41.5;][]{Bertin2002} to place the images into a common grid of $5250\times5250$ pixels, using as resampling method LANCZOS3.

\subsubsection{Sky subtraction}\label{sect:sky_subtraction}

Subtracting the sky is a crucial step that has to be tackled carefully in order not to introduce artificial structures or produce oversubtraction. We perform the sky subtraction individually on each frame, and, since the field-of-view of the data is not very large, we can work under the assumption of constant background. Additionally, not assuming any structure for the background rules out the possibility of introducing any artificial structure in the data.

To characterise the background, we build a mask that accounts for as much signal as possible. For this, we first build an aggressive mask using \texttt{NoiseChisel} and later we manually apply a mask centered on the galaxy with a radius of 120\arcsec{}. This mask over the galaxy is large enough so any region of undetected faint structure in its outskirts is not taken into account for the sky estimation. With the remaining unmasked pixels, we compute the sky value using a sigma clipping median and subtract it from each of the frames.

\subsubsection{Data combination}

Once we astrometrize the individual frames and subtract their sky, they are ready to be combined. We combine the frames using a weighted mean, where each of the frames is weighed based on the standard deviation of its background. The reason behind this weighing scheme is that the quality of the frames varies between nights and even within one single night (e.g., illumination conditions, airmass, gradients related to the orientation of the observation), so this combination allows us to value more data of higher quality. Additionally, to perform an optimal combination, we apply the weights using a quadratic scheme, with the weight of the $i^{th}$ frame $\sigma_{min}^{2} /\sigma_{i}^{2}$, where $\sigma_{min}$ corresponds to the standard deviation of the less noisy frame.\\

\begin{figure*}[!htb]
    \centering
    \includegraphics[width=\linewidth]{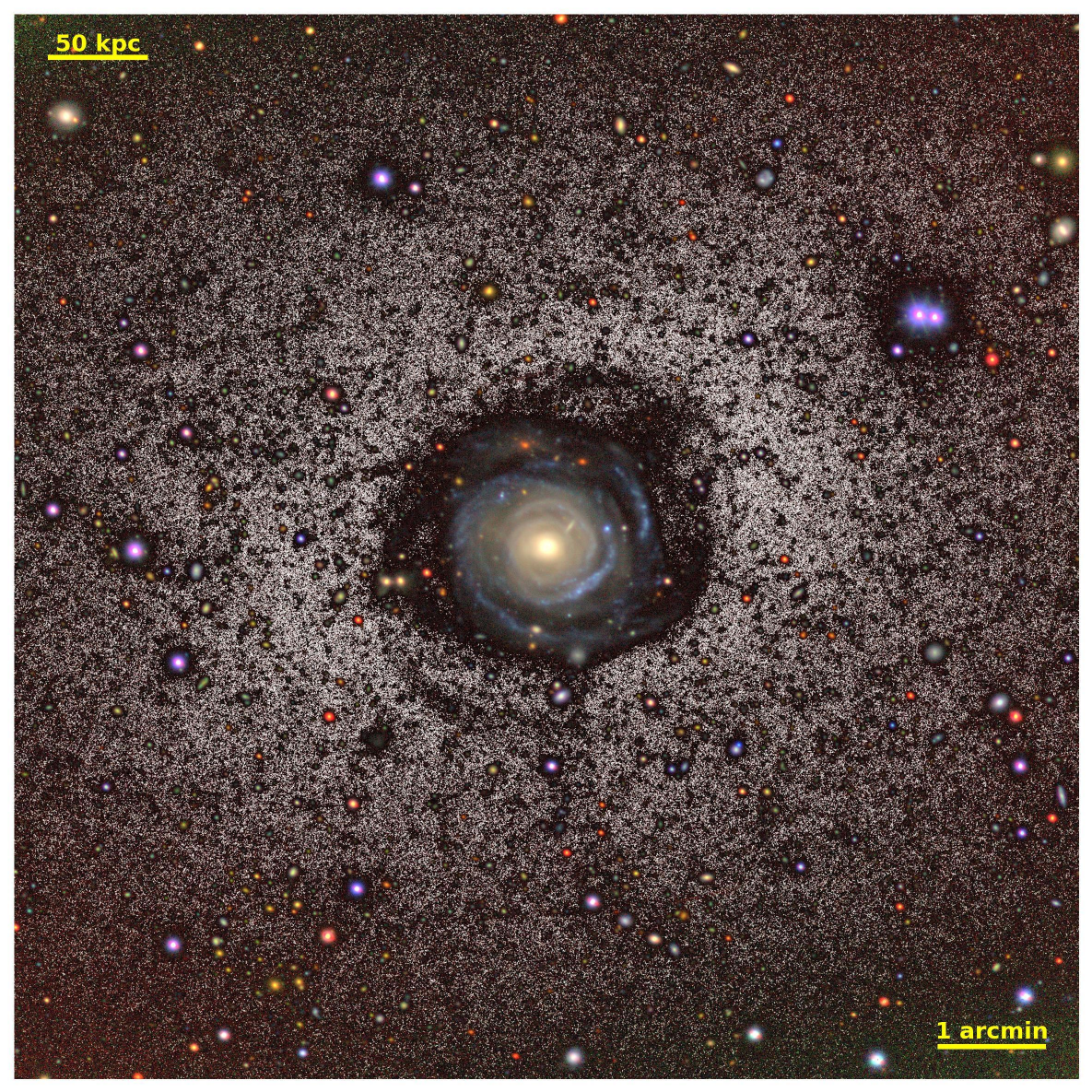}
    \caption{Deep image of Malin~2 obtained from the Two-meter Twin Telescope (TTT) at Teide Observatory. The color composite image is created using \textit{g}, \textit{r}, and \textit{i}-band filters with the \texttt{Gnuastro} package \texttt{astscript-color-faint-gray} \citep{Infante-Sainz2024}. The black and white background corresponds to the \textit{g}-band image. The image covers a field of view of $10\arcmin{} \times 10\arcmin{}$ or $542.4 \mathrm{kpc} \times 542.4 \mathrm{kpc}$. The North is up and the West is right.}
    \label{fig:Malin2_full_field_color}
\end{figure*}

The stacked image is deeper than the individual exposures, allowing faint structure that is below the noise in the individual images to now emerge. This previously undetected signal biased our estimation of the background (making us overestimate it), so we can now seize the information from the stacked image to define a more complete mask and apply it to the individual frames. With this improved mask, we calculate again the background estimation and go over all the subsequent steps described above, thus obtaining a final (non-calibrated) stacked image as shown in Fig. \ref{fig:Malin2_full_field_color}.

\subsubsection{Photometric calibration}

The last step is to calibrate the stacked image. We convert pixel values from ADUs to physical units (Janskys) by matching the fluxes of stars in our image to the fluxes of stars in an already-calibrated survey. The survey used for the calibration is the Panoramic Survey Telescope and Rapid Response System (Pan-STARRS) DR1\citep{panstarrs}. 

We download the Pan-STARRS field of Malin~2 and generate a catalogue of sources. We then match this catalogue with Gaia DR3 stars to obtain a clean sample of bona fide stars. Since we want to maximize the number of stars used for the calibration, instead of using just stars identified in Gaia, we look at the range of FWHM that these stars have, using it as a criterion for selecting point-like sources in the Pan-STARRS images. Then we perform the photometry of the selected sources by using large enough apertures, which ensures that we virtually measure all the flux of the point-like source. If the aperture were too small, some flux would be lost in the tails of the PSF, generating inconsistencies when comparing fluxes measured on different images due to the differences in the PSFs. The use of a large number of stars effectively washes out the effect of the contaminants introduced by the large aperture. To define the aperture to use, we explore a wide range of apertures, studying at which value the measured flux stabilizes. We conclude that for Pan-STARRS data, apertures of $r=9R_e$ are optimal. We then apply the same process to the data to calibrate, obtaining a catalog of point-like sources to be used for the calibration. The apertures used for measuring the flux are $r=7R_e$. Additionally, to use non-saturated stars with high SNR, we use for the calibration only sources with a magnitude between a certain range (15.5 < mag < 19.0, 15.0 < mag < 18.0, and 14.5 < mag < 17.0 for the \textit{g}-, \textit{r}- and \textit{i}-bands, respectively).

Since the photometric calibration is a complex task, surveys have non-negligible offsets between them. Thus, in order to be able to fairly compare data calibrated with different surveys, we decide to reference our calibration to an arbitrary but common framework, this being the Gaia spectra. This introduces a correction that we apply to the fluxes measured in Pan-STARRS, which are compared with the fluxes measured from the Gaia spectra and corrected (by applying an offset) to match them. Secondly, in order to give a precise photometry, we need to take into account the transmittance of the filters involved in the calibration. Even if technically we are working with the same filters (in this case $g, r, i$ from Pan-STARRS and from TTT), there are differences in the filter response between the filters of different telescopes. This can be alleviated by applying a color correction. We characterize the applied color correction based on the $g-r$ color and is determined using the Pan-STARRS filter shapes, the shapes of our filter, and GAIA spectra from stars in the field.

Finally, we match the catalogues and calculate the factor that we need to apply to our image in order to have it calibrated to physical units. The zero-point has been set to 22.5 in the AB magnitude system.

%%%%%%%%%%%%%%%%%%%%%%%%%%%%%%%%%%%%%%%%%%%%%%%%%%%%%%%%%%%%%%
\section{Analysis and Results} \label{sect:results}

\subsection{Discovery of new low surface brightness features in Malin~2}\label{sect:discovery_of_lsb_features}

Our deep imaging of Malin~2 reveals, for the first time, several previously unseen low surface brightness structures in the extended disk of the galaxy as shown in Fig. \ref{fig:Malin2_full_field_color}. In the northwest region, we detect a broad, diffuse stellar emission extending out to nearly 100\,kpc from the galaxy center. This diffuse emission appears to align well with the outer \Hi{} contours from \citet{Pickering1997} (see Sect. \ref{sect:hi_distribution} for more details). In the southeast of the galaxy, we see an elongated, spiral arm-like structure emerging beyond the main optical disk, reaching surface brightness levels of $\mu_g \approx 30$\,\magperarcsec{} at radii of $\sim95$\,kpc. These features were barely detected and never reported in any of the previous imaging of Malin~2 (e.g., SDSS, DECaLS) due to the limited surface brightness sensitivity of these surveys. Fig. \ref{appendix:figure_sdss_decals_ttt} shows a comparison of the SDSS, DECaLS, and TTT imaging of Malin~2.

Our newly confirmed LSB structures show distinct morphological characteristics. The northwest diffuse emission exhibits a rather asymmetric structure with no clear spiral pattern, while the southeastern feature displays an arm-like morphology similar to an outer spiral structure or a tidal feature. These features demonstrate that Malin~2's disk extends well beyond previous estimates and exhibits complex, non-axisymmetric structure in its outer regions.

Additionally, in the southeast region, at a projected distance of $\sim$130\,kpc from the galaxy center, we identify a very diffuse source that appears to be a satellite galaxy of Malin~2 that is previously unknown. We provide a detailed photometric analysis of this source in Sect. \ref{sect:satellite}.

\subsection{Surface brightness profile measurements}\label{sect:sb_profile_measurements}

\begin{figure*}[htb]
  \centering
  \begin{minipage}[c]{0.48\textwidth}
    \centering
    \vspace{0.02\textheight}
    \includegraphics[width=\linewidth]{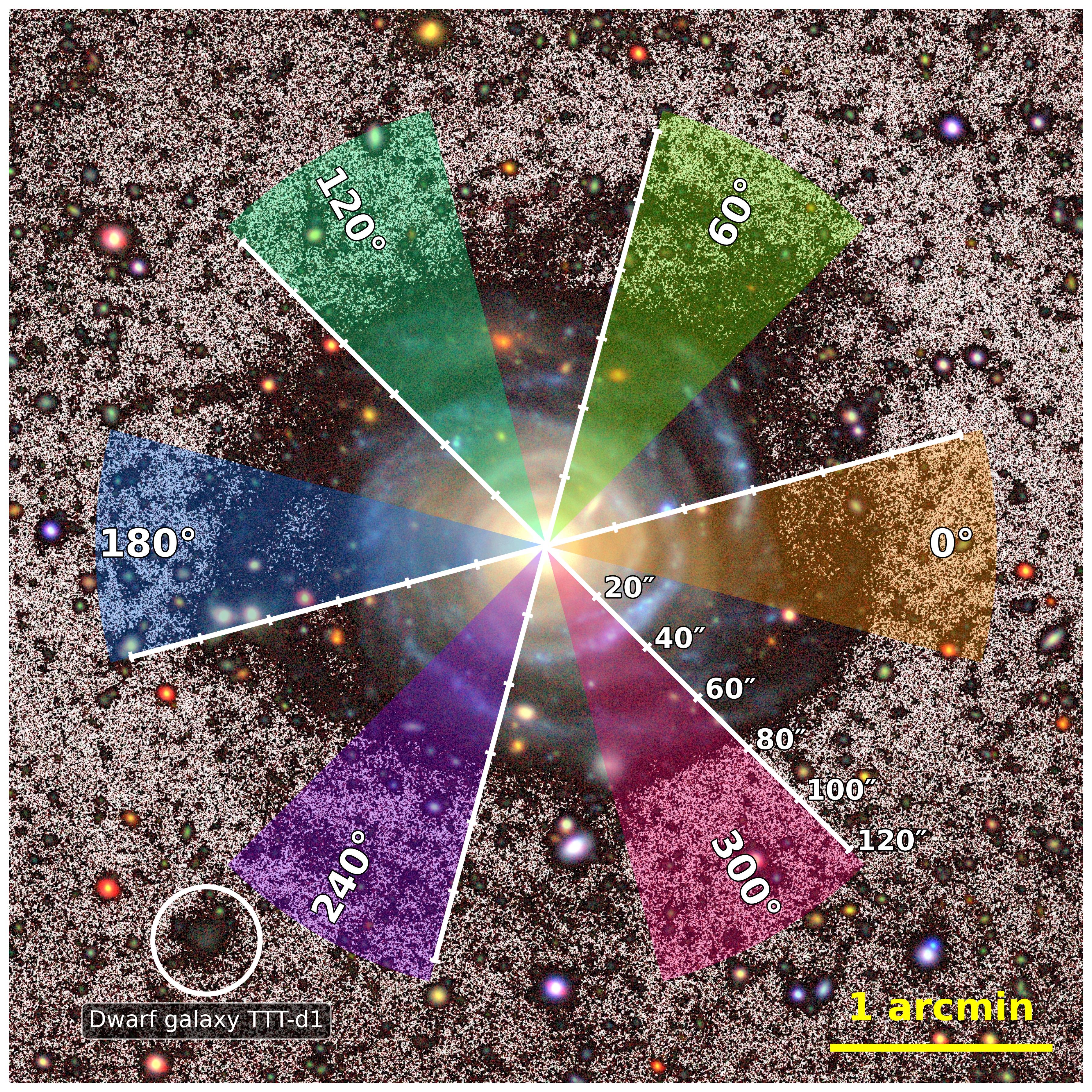}
  \end{minipage}%
  \hfill
  \begin{minipage}[c]{0.48\textwidth}
    \centering
    \includegraphics[width=\linewidth]{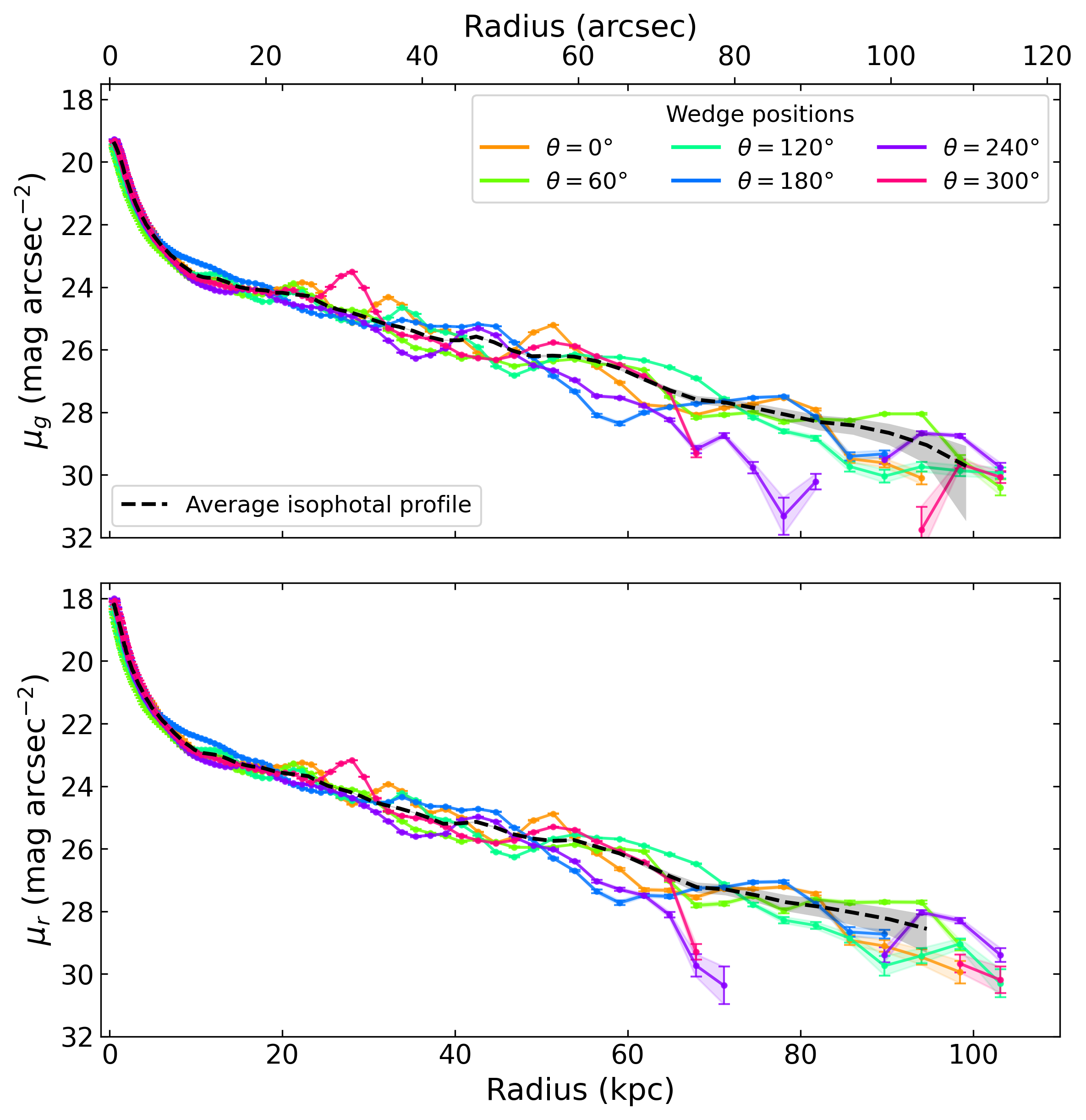}
  \end{minipage}
  \caption{Multi-directional photometry of Malin~2. Left: Wedge positions used for the photometry overplotted on the color image of Malin~2. Six wedges, each with 30° opening angles are positioned at 0°, 60°, 120°, 180°, 240°, and 300°. A radial scalebar (in units of arcsec) is also shown along each wedge for easier comparison with the radial profiles. The location of the candidate dwarf satellite galaxy TTT-d1, discussed in Sect. \ref{sect:satellite}, is marked in the bottom left corner (white circle). Right: Corresponding radial profiles of $g$- and $r$-band surface brightness for each wedge direction, showing significant structural asymmetry across different orientations. The black dashed line is the classical azimuthally averaged isophotal profile we measure, with the gray shaded region as its uncertainty.}
  \label{fig:malin2_photometry}
\end{figure*}

The complex and asymmetric morphology of Malin~2, visible in our deep imaging, makes a traditional, azimuthally averaged surface brightness profile inadequate for characterizing its structure. Therefore, we perform a multi-directional radial profile measurement using wedge-shaped sectors to probe different galaxy orientations and quantify the structural asymmetries. We place six wedges at angles of 0°, 60°, 120°, 180°, 240°, and 300° (measured counter-clockwise from west towards north), each with an opening angle of 30° (see Fig.~\ref{fig:malin2_photometry}, left panel).

Before making radial profile measurements along the wedges, we create a mask to remove any background/foreground contaminants of the photometry. We create the mask using the \texttt{MTObjects} tool \citep{Teeninga2016}, on the combined $g+r$ image. We run \texttt{MTObjects} using a $\text{\texttt{move\_factor}} = 0.3$, optimized for the detection of faint structures. We later visually inspect and manually edit the mask when necessary to ensure that all the contaminants are properly masked and none of the structures of Malin~2 are masked (e.g., spiral arms, diffuse regions). Hereupon, we use this mask throughout for our multi-band photometry.

Within each wedge shown in the left panel of Fig.~\ref{fig:malin2_photometry}, we extract surface brightness profiles with logarithmic radial bins ranging from the center of the galaxy to $\sim$125\,kpc (140\arcsec{}) until we see no visible signal in the optical images. We measure the local background sky level and noise at the last radial bin for each wedge separately. This is to take into account the local variations in the sky in addition to the global background subtraction we performed in Sect. \ref{sect:sky_subtraction}. We truncate the radial profiles along each wedge at the radius where they reach the background noise. We perform the above wedge profile measurements for both the \textit{g}- and \textit{r}-band images. We do not attempt to do a similar procedure for our \textit{i}-band data, as it is about 2 magnitudes shallower than our deepest \textit{g}-band data (see Sect. \ref{sect:data}). The \textit{i}-band data in this work is used only to create the color images. From hereupon, we correct all the radial profiles for foreground Galactic extinction using the \citet{Schlegel1998} dust maps normalized by \citet{Schlafly2011} and a \citet{cardelli1989} dust extinction law. We also correct all the profiles for inclination following \citet{Trujillo2020}\footnote{For the inclination correction, we assume a ratio of scale height to scale length of $z_0/h = 0.12$ as given in Table 1 and Eq. 2 of \citet{Trujillo2020}.}. For the inclination correction, we estimate the axis ratio of Malin~2 by running the \texttt{AutoProf} tool \citep{Stone2021} on our \textit{g}-band image, which is the deepest. We obtain an axis of $b/a = 0.91$ (inclination $i=23.9\degree$) and a position angle (PA) of $72.7\degree$ corresponding to the 26\,\magperarcsec{} isophote in \textit{g}-band from the \texttt{AutoProf} fit. This value of inclination is about $14\degree$ lower than previously reported for Malin~2, whereas the PA is well consistent with the previous estimates \citep[e.g., ][]{Pickering1997,Kasparova2014}. However, since we have deeper imaging of Malin~2 in this work, we choose to adopt the values we estimate and correct all our radial profiles for inclination throughout this work.

Figure~\ref{fig:malin2_photometry} (right panel) presents the $g$- and $r$-band surface brightness profiles we measure for each wedge. The profiles reveal significant variations along different directions, with surface brightness differences of up to 2~mag\,arcsec$^{-2}$ at large radii ($R>60$\,kpc) between different wedges. Most of the profiles reach down to a surface brightness level of $\sim 30$\,\magperarcsec{} and a radial extent of $\sim110$\,kpc in both the bands. We verify that our profiles are consistent with the SDSS $g$- and $r$-band azimuthally averaged profiles from \citet{Kasparova2014}. However, their profiles only reach a surface brightness level of 26\,\magperarcsec{} (up to a 55\,kpc radius), while ours extend about 4 magnitudes deeper and to twice the radial extent. The 240° and 300° wedges in our profiles show some early drop around 65\,kpc compared to the other directions, as visible from the optical image. However, we see some additional stellar emission along these wedges at a radial range of $90-100$\,kpc, with a \textit{g}-band surface brightness of $\sim 29$\,\magperarcsec{}. This corresponds to the extended spiral arm-like structure we observe in the south of the galaxy, as discussed in Sect. \ref{sect:discovery_of_lsb_features}. Similarly, along the 60° wedge, we see a clear excess in stellar emission within $80-100$\,kpc radius and nearly constant \textit{g}-band surface brightness of $\sim 28$\,\magperarcsec{}. This corresponds to the diffuse stellar emission we see in the north-east part of the galaxy, as mentioned in Sect. \ref{sect:discovery_of_lsb_features}. The 180° wedge also shows a similar trend where we see an almost flat \textit{g}-band surface brightness of $\sim 28$\,\magperarcsec{} within the radial range of $60-80$\,kpc, corresponging to the outer diffuse emission we see along this wedge in the left panel of Figure~\ref{fig:malin2_photometry}. Apart from the extended disk of the galaxy, in the inner regions ($R<60$\,kpc), the radial profiles along the different wedges seem to follow a similar trend on average, except for the visible local wiggles in the profiles that are likely due to the spiral arms (for instance, in the 300° wedge profile, we can clearly see a bump around 30\,kpc that is due to the bright star forming spiral arm visible in the optical image at this radius).

For a comparison to the wedge profiles, in Fig.~\ref{fig:malin2_photometry}, we also show the traditional azimuthally averaged surface brightness profiles of Malin~2. We obtain these average profiles by placing elliptical isophotes at a fixed position angle and axis ratio ($\mathrm{PA}=72.7\degree$, $b/a = 0.91$) using the {\tt Isophote\_Extract} method of the {\tt AutoProf} tool \citep{Stone2021}. The profiles are extracted using the same mask as used for the wedge profiles. We measure the background estimate and noise along each isophote by placing 20 square boxes each of size 30\arcsec{} at a distance of 140\arcsec{} around the galaxy, following \citet{GildePaz2005} and \citet{Junais2022}. We can see that the average profiles shown in Fig.~\ref{fig:malin2_photometry} agree well with the wedge profiles, but the directional asymmetries we see in the wedge profiles are mostly washed out, as expected. This further illustrates the need for our wedge-based approach. Therefore, throughout this work, we focus more on the wedge profiles rather than the average profiles.

\subsection{Color and stellar mass surface density profiles}\label{sect:color_stellar_mass_surface_density}

\begin{figure}[!htb]
    \centering
    \includegraphics[width=\linewidth]{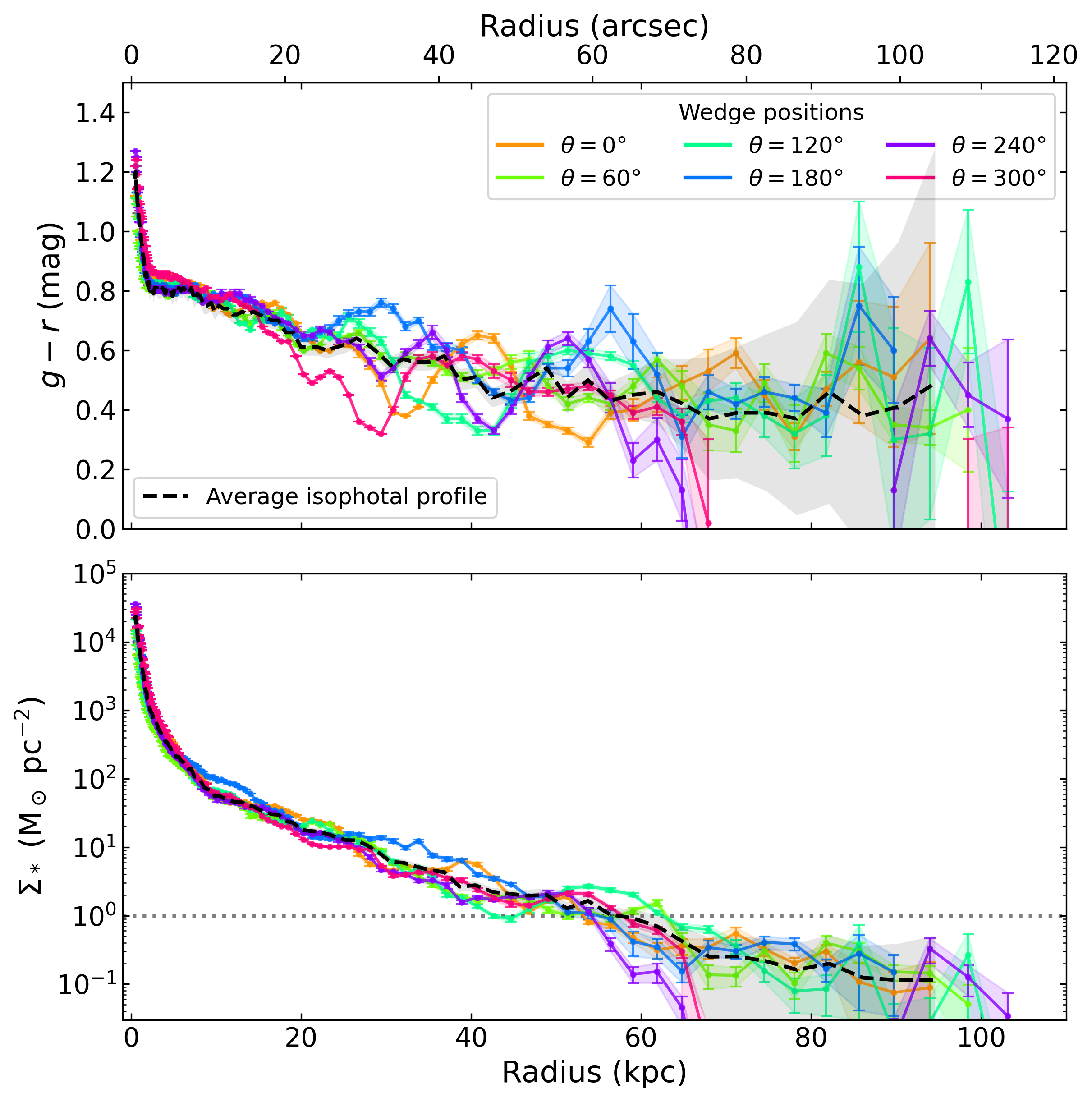}
    \caption{Radial profiles of $g-r$ color (top) and stellar mass surface density $\Sigma_*$ (bottom) for each wedge direction in Malin~2. The horizontal dotted line in the bottom panel marks 1~\msunpcsq{} stellar mass surface density, where we estimate the $R_1$ radius as discussed in Sect. \ref{sect:r1_radii}. The colored legends for the wedge profiles are the same as in Fig. \ref{fig:malin2_photometry}. The black dashed line is the classical azimuthally averaged isophotal profile we measure, with the gray shaded region as its uncertainty.}
    \label{fig:malin2_color_density}
\end{figure}

Figure~\ref{fig:malin2_color_density} shows the $g-r$ color profiles (top panel) and the stellar mass surface density profiles (bottom panel) of Malin~2. We compute the stellar mass surface density profiles using Eq. 1 of \citet{Bakos2008}, with a combination of the $g-r$ color and \textit{g}-band surface brightness profiles, following the color mass-to-light ratio relation from \citet{Roediger2015}. Similar to the surface brightness profiles, both the color and stellar mass surface density profiles show diverse behavior across different directions. Colors vary from red to blue with $g-r \sim 1.2$~mag in the center of the galaxy to $\sim 0.4$~mag in the extended disk (in the case of the 240° and 300° wedges, $g-r$ reaches up to $\sim0.2$~mag). Within the radial range of 20\,kpc to 60\,kpc, the color shows significant fluctuations along the different wedges due to the presence of several spiral arms. For instance, the 300° wedge profile shows a big drop towards a bluer color around 30\,kpc radius as it passes through a bright spiral arm (similar to what we see in the surface brightness profiles and the optical image from Fig. \ref{fig:malin2_photometry}). In contrast, the 180° wedge shows two red peaks around 30\,kpc and 56\,kpc radii, which also appear in the $g-r$ color map shown in Fig.~\ref{appendix:figure_2d_g-r_map}, where we observe extended red emission along these radii. We confirm that these red peaks correspond to stellar populations in Malin~2 and are not due to any contamination from background emission.

Beyond the radius of 60\,kpc, the 240° and 300° wedges show an early drop around 65\,kpc, whereas the rest of the wedges extend well until a radius of $\sim100$\,kpc (note that in the 240° and 300° wedges also we see some localized emission around 100\,kpc, corresponding to the extended southeastern spiral-like structure we see in the optical images). In the extended disk of the galaxy ($R > 60$\,kpc), the color profiles of all the wedges mostly stay flat around $g-r = 0.4$ mag, with a possible reddening beyond 90\,kpc, however, with a large scatter. Apart from the one-dimensional radial color profiles shown in Fig. \ref{fig:malin2_color_density}, the $g-r$ color map shown in Appendix \ref{appendix:2d_g-r_map} shows the overall color distribution in Malin~2, both on and off the wedges we use in our measurements.

Figure~\ref{fig:malin2_color_density} (bottom panel) shows the stellar mass surface density profile of Malin~2 along the wedges. It follows a rather smooth trend in all the wedges. The stellar mass surface density of the profiles reaches a peak value of $\sim4\times10^4$~\msunpcsq{} in the center, with a steep decline within the 20\,kpc by about 3 orders of magnitude, and reaches down to $\sim0.1$~\msunpcsq{} at the outermost part of the disk around 100\,kpc. All the wedges follow a similar trend up to 60\,kpc radius (with minor differences corresponding to the spiral arms along the wedges). Similar to what we see in the color profiles, the stellar mass surface density profiles of the 240° and 300° wedges show a steep drop around 65\,kpc radius. This is likely the optical edge of the galaxy at these azimuths, as we do not see any significant asymmetries along these two wedges.
The remaining wedges all show a rather constant density value of $\sim0.3$\,\msunpcsq{} beyond this radius (see Table \ref{table:wedge_statistics}). Figure~\ref{fig:malin2_color_density} thus shows that the outer diffuse asymmetrical structures we see in the optical images within the radial range of 60 to 100\,kpc (beyond the main optical disk) have, on average, similar stellar density and colors, indicating a likely common origin for these structures.

\subsection{Size estimates and stellar population properties of the extended disk} \label{sect:r1_radii}

\begin{table*}[!htb]
    \centering
    \caption{Photometric properties of Malin~2 along different wedge directions.}
    \begin{tabular}{c c|cccc}
    \hline
    \hline
    Wedge Angle & $R_1$ & \multicolumn{4}{c}{Mean photometry of the extended disk at $R > 70$\,kpc} \\
    \cline{3-6}
    (deg) & (kpc) & $\langle \mu_g \rangle$ & $\langle \mu_r \rangle$ & $\langle g - r \rangle$ & $\langle \Sigma_\star \rangle$ \\
    &   & (mag arcsec$^{-2}$) & (mag arcsec$^{-2}$) & (mag) & (M$_\odot$ pc$^{-2}$) \\
    \hline
    0   & 53.4 $\pm$ 0.2 & 28.60 $\pm$ 0.04 & 28.10 $\pm$ 0.05 & 0.50 $\pm$ 0.07 & 0.24 $\pm$ 0.03 \\
60  & 63.3 $\pm$ 0.4 & 28.30 $\pm$ 0.02 & 27.88 $\pm$ 0.03 & 0.42 $\pm$ 0.03 & 0.20 $\pm$ 0.02 \\
120 & 62.6 $\pm$ 0.5 & 29.16 $\pm$ 0.05 & 28.78 $\pm$ 0.07 & 0.39 $\pm$ 0.09 & 0.15 $\pm$ 0.05 \\
180 & 54.9 $\pm$ 2.6 & 28.26 $\pm$ 0.03 & 27.75 $\pm$ 0.04 & 0.51 $\pm$ 0.05 & 0.28 $\pm$ 0.05 \\
240 & 54.3 $\pm$ 0.5 & 29.18 $\pm$ 0.05 & 28.78 $\pm$ 0.09 & 0.40 $\pm$ 0.10 & 0.13 $\pm$ 0.04 \\
300 & 57.9 $\pm$ 0.4 & -- & -- & -- & -- \\
\hline
    \end{tabular}
    \tablefoot{(1) Wedge angle; (2) $R_1$ radius corresponding to the radius at which the stellar mass surface density reaches 1\,\msunpcsq{}; (3 - 6) The mean photometry of the extended disk between a radial range of 70\,kpc to about 110\,kpc. The columns mark the mean \textit{g}- and \textit{r}-band surface brightness levels, mean $g-r$ color, and the mean stellar mass surface density, respectively.}
    \label{table:wedge_statistics}
    \end{table*}

We measure the extent of the stellar disk in each wedge using the $R_1$ radius, defined as the radius where the stellar mass surface density falls to 1\,\msunpcsq{} \citep{Trujillo2020}. \citet{Trujillo2020} chose that stellar mass surface density of 1\,\msunpcsq{} as that roughly corresponds to a proxy for in-situ star formation threshold in galaxies \citep[e.g.,][]{Schaye2004,Martinez-Lombilla2019}. For galaxies with the stellar mass of the Milky Way, $R_1$ is a good proxy for the location of the edge of the galaxy \citep{Chamba2022}. Recent works show that the $R_1$ radius (a proxy for the edge) provides a physically motivated measure of galaxy size based on the end of the in-situ star formation rather than the traditional scale length, effective radius, or $R_{25}$ values \citep{Trujillo2020,Chamba2020,Chamba2022}. 

Table~\ref{table:wedge_statistics} shows our measured $R_1$ radius for the different wedge profiles. We adopt the last radius at which the stellar mass surface density profiles that pass through the threshold of 1\,\msunpcsq{}. For instance, the 60° wedge profile passes through the value of 1\,\msunpcsq{} twice, around 45\,kpc and 63.3\,kpc radii. We adopt the latter value. The $R_1$ values range from a minimum of $\sim53$\,kpc in the 0° wedge to $\sim63$\,kpc in the 60° direction. The northern wedges (60° and 120°) show larger $R_1$ values, above 60\,kpc, consistent with the extended low surface brightness features we detect in this region (see Fig. \ref{fig:malin2_photometry}). All the other wedges have similar $R_1$ values around 55\,kpc. From all the wedge profiles, we get a mean $R_1$ radius of $57.7$\,kpc for Malin~2. Note that the true extent of Malin~2 at lower surface brightness is even beyond this radius, as we can see from our profiles that it reaches up to a radius of about 110\,kpc. On the other hand, it is worth noting that the $R_1$ coincides pretty well with the location of the edge of the disk (i.e., where the mass profile drops in those wedges, non-affected by the tidal-like features).

We also estimate the effective radius ($R_{e}$) of Malin~2 using the average isophotal \textit{g}-band surface brightness profile shown in Fig. \ref{fig:malin2_photometry} (since by definition $R_{e}$ depends on the total light of the galaxy, it is only meaningful to use the average profile for the $R_{e}$ estimate than the wedge profiles). We estimate $R_{e}$ by non-parametrically integrating the average \textit{g}-band profile to find the radius corresponding to the half of the total measured light. We obtain an $R_{e}$ of 20.3\,kpc. This is about 2.8 times smaller than our mean $R_1$ radius of 57.7\,kpc. The mean $R_1$ is also 2.5 times larger than the \textit{g}-band disk scale length of $23.5\pm4.5$\,kpc ($26\pm5$ arcsec) reported by \citet{Kasparova2014} for Malin~2 from a bulge-disk decomposition of an azimuthally averaged profile. Due to the asymmetric nature of our profiles, in this work, we do not attempt to perform a similar bulge-disk decomposition. A detailed comparison of our measured $R_1$ radii with other spiral galaxies from the literature is given in Sect. \ref{sect:comparison_to_other_spirals}.

Table~\ref{table:wedge_statistics} also shows our measured mean stellar properties of the extended disk of Malin~2 beyond a radius of 70\,kpc. We adopt the value of $R > 70$\,kpc to probe the diffuse stellar population visible in our imaging beyond the $R_1$ radius. We verify that a choice of a different value (e.g., $R > 60$\,kpc) for estimating the mean stellar properties in the outer disk does not make a significant change in the trends. All the wedges in this radial range, except the 300° wedge with a non-detection, show similar surface brightness level, color, and stellar mass surface density. The \textit{g}-band surface brightness in these wedges (0°, 60°, 120°, 180° and 240°) falls within a range of 28.3 to 29.2\,\magperarcsec{} (a similar trend is seen in the \textit{r}-band surface brightness levels too with a range of 27.8 to 28.8\,\magperarcsec{}). This gives a $g-r$ color in the range of 0.39 to 0.51 mag and a stellar mass surface density between 0.13 and 0.28\,\msunpcsq{}. 
Therefore, the extended disk of Malin~2 has a mean $g-r$ color of 0.44 mag and stellar mass surface density of 0.2\,\msunpcsq{}. This indicates a relatively blue and young stellar population. Moreover, a rather similar stellar population across these multiple wedges points to a common origin for the diffuse emission in the extended disk of Malin~2. As an alternate approach to the wedge-based photometry, we also perform aperture photometry on several visually selected regions on the extended disk of Malin~2. We provide the results of these measurements in Appendix \ref{appendix:aperture_photometry_regions}. We find that they are consistent with our measurements from the wedge profiles.

\subsection{Discovery of a diffuse galaxy: A likely new satellite of Malin~2}\label{sect:satellite}

\begin{figure}[!htb]
    \centering
    \includegraphics[width=\linewidth]{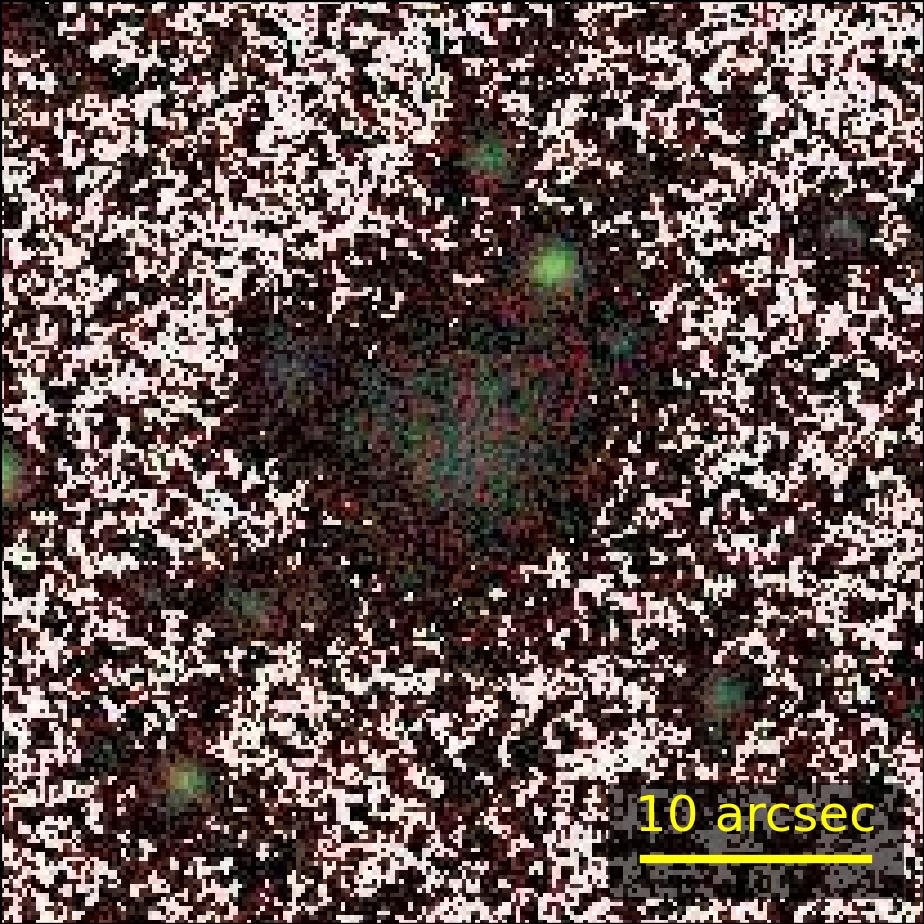}
    \caption{Dwarf galaxy TTT-d1 discovered in the vicinity of Malin~2. The color composite image is created using \textit{g}, \textit{r}, and \textit{i}-band filters with the \texttt{Gnuastro} package \texttt{astscript-color-faint-gray} \citep{Infante-Sainz2024}. The black and white background corresponds to the \textit{g}-band image. The image covers a field of view of $40\arcsec{} \times 40\arcsec{}$ or $36.16 \mathrm{kpc} \times 36.16 \mathrm{kpc}$.}
    \label{fig:malin2_dwarf}
\end{figure}

We identify a diffuse, LSB object that appears to be a satellite galaxy of Malin~2, in the southeast outskirts at a projected distance of 2.43\arcmin{} (131.58\,kpc at the distance of Malin~2) from the center of Malin~2, close to the edge of the 240° wedge in the left panel of Fig. \ref{fig:malin2_photometry}. Here we report the discovery of this galaxy, clearly visible in our deep imaging, but barely visible in previous surveys (see Fig. \ref{fig:malin2_dwarf}). We name the dwarf galaxy TTT-d1 (the first dwarf galaxy discovered by the TTT telescope). TTT-d1 is located at coordinates of R.A,  DEC: 10h39m 59.26s, +20d48m59.15s.

To quantify the photometric and morphological properties of TTT-d1, initially we perform a {\tt GALFIT} \citep{Peng2010} S\'ersic modeling of the \textit{g}- and \textit{r}-band images of the galaxy. We create the masks for the {\tt GALFIT} modeling using the procedure discussed in Sect. \ref{sect:sb_profile_measurements}. We present the results of our {\tt GALFIT} S\'ersic modeling in Table \ref{appendix_table:galfit_results}. We find that TTT-d1 has a \textit{g}-band effective radius of $6.38\pm0.22$\arcsec{}, S\'ersic index (\textit{n}) of $0.86\pm0.03$ and a central surface brightness ($\mu_{0,g}$) of $25.94\pm0.09$\,\magperarcsec{}. We find similar quantities in the \textit{r}-band, with a slightly smaller effective radius ($5.88\pm0.23$\arcsec{}). We obtain a mean $g-r$ color of $0.53\pm0.06$\,mag for TTT-d1, indicating a redder stellar population. 
We extract azimuthally averaged radial surface brightness profiles for TTT-d1 in the \textit{g}- and \textit{r}-bands using the {\tt Photutils} python package \citep{photutils}. We use the same mask as in the previous step and follow the surface brightness profile measurement procedures from \citet{Junais2022}. Figure \ref{appendix_fig:dwarf_profiles} shows our extracted profiles, overplotted with the best-fit {\tt GALFIT} models obtained in the previous step. We can clearly see that the measured radial profiles of TTT-d1 are consistent with our independently estimated single S\'ersic profiles.

Assuming the same distance of Malin~2, TTT-d1 has a \textit{g}-band effective radius of $5.77\pm0.20$\,kpc, central surface brightness of $\sim$ 26\,\magperarcsec{} and a stellar mass\footnote{We estimate the stellar mass of TTT-d1 by converting the \textit{g}- and \textit{r}-band profiles to stellar mass surface density as described in Sect. \ref{sect:color_stellar_mass_surface_density}. Assuming the distance of Malin~2, the stellar mass surface density profiles were integrated until the last measured radius to obtain the total stellar mass.} of $2.11\times10^{8}$ \msun{}. These properties are consistent with the known population of ultra-diffuse galaxies (UDGs) and demonstrate that such objects can exist as satellites in the halos of GLSB galaxies. With its large radial extend of $\sim$ 6\,kpc, TTT-d1 will also be among the largest known UDG and comparable to the Nube galaxy from \citet{Montes2024}. However, we should be cautious with such an inference, as we do not yet have an independent distance estimate for this source to confirm its association with Malin~2. While the large size of the TTT-d1 may reduce the likelihood of it being at the same distance as Malin2, if confirmed, it could offer valuable constraints on the formation and evolution of GLSB galaxies. This will be done in future work.

%%%%%%%%%%%%%%%%%%%%%%%%%%%%%%%%%%%%%%%%%%%%%%%%%%%%%%%%%%%%%%
\section{Discussion} \label{sect:discussion}

\subsection{Comparison of Malin~2 to other spiral galaxies}\label{sect:comparison_to_other_spirals}

\begin{figure}[!htb]
  \centering
  \includegraphics[width=0.48\textwidth]{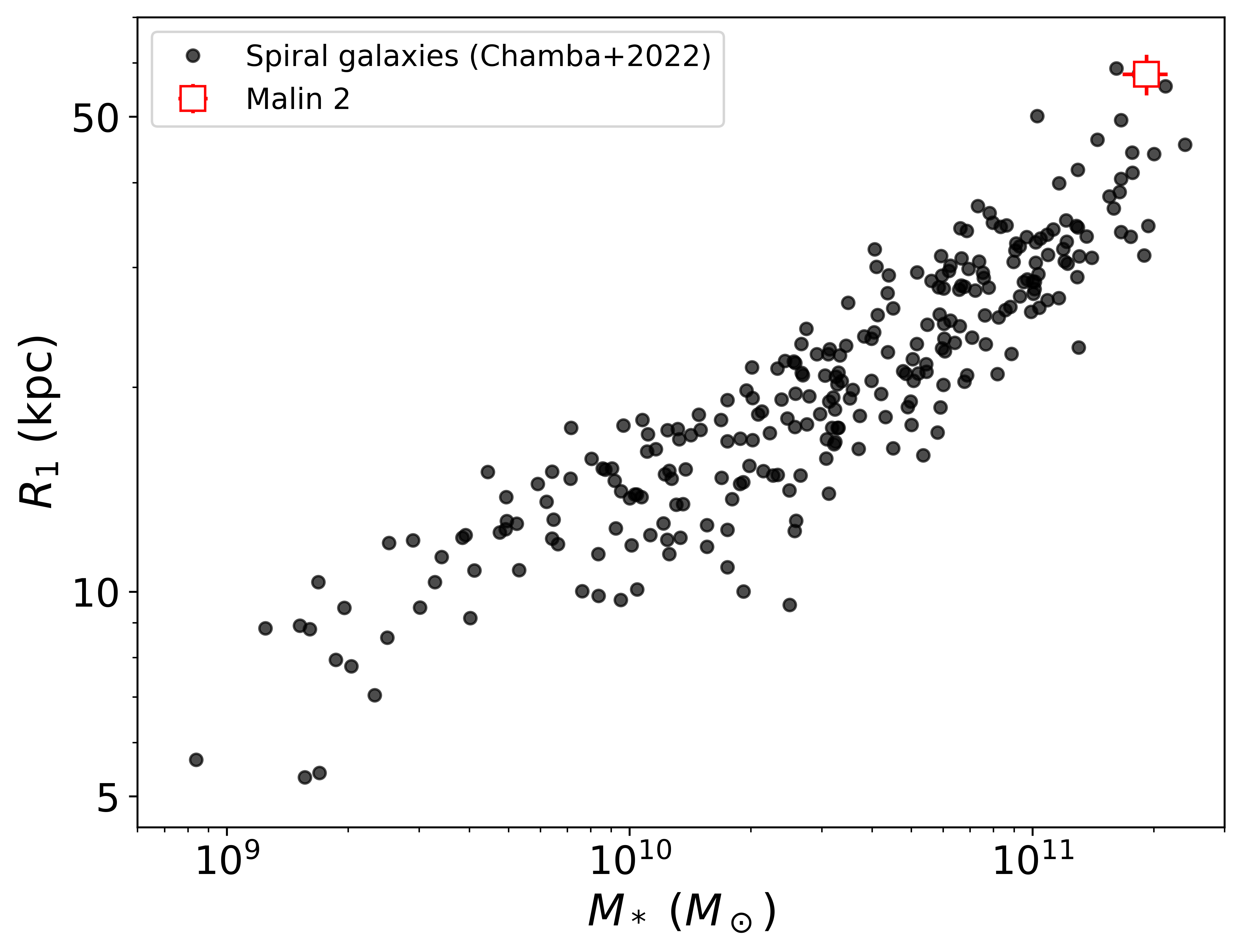}
  \caption{Stellar mass versus $R_1$ radius relation. Malin~2 (red open square) is compared to other spiral galaxies at $z<0.1$ from \citet{Chamba2022} (black circles). Malin~2 clearly stands out as one of the extreme cases in this relation.}
  \label{fig:mstar_r1}
\end{figure}

To place Malin~2 in the context of the broader galaxy population, we compare its properties with a large sample of 273 spiral galaxies\footnote{As the original sample of \citet{Chamba2022} also contains elliptical galaxies, here we include only the spiral galaxies with a TType $> -1$ from their sample.} at $z<0.1$ from \citet{Chamba2022}. Figure~\ref{fig:mstar_r1} shows Malin~2's position in the stellar mass versus $R_1$ radius relation. With a stellar mass\footnote{We estimate the stellar mass of Malin~2 by integrating the stellar mass surface density profiles shown in Fig. \ref{fig:malin2_color_density} for the different wedges. We use the mean stellar mass obtained from these multi-wedge profiles.} of $\log(M_*/M_\odot) = 11.3 \pm 0.1$ and a mean $R_1$ radius of $57.7\pm4.0$\,kpc, Malin~2 clearly stands out as one of the extreme cases, lying along the most massive and extended regime of spiral galaxies. This confirms Malin~2 as one of the most extended spiral galaxies known to date. 

The extreme size of Malin~2 can be visually illustrated by comparing it with scaled images of a few other spiral galaxies. Figure~\ref{fig:malin2_comparison_full} (top panels) shows the image of Malin~2 compared to that of a spiral galaxy of similar stellar mass (J005042.73+002558.37; $\log(M_*/M_\odot) = 11.2$) and a Milky Way-mass galaxy (J012015.34-002009.00; $\log(M_*/M_\odot) = 10.6$), all scaled to the same physical dimensions in kpc. The Malin~2 image spans 271\,kpc in width, whereas the other two images are 134\,kpc and 69\,kpc, respectively. We can see a significant difference in their radial extent. The Milky Way-mass galaxy (J012015.34-002009.00) appears significantly smaller in comparison. The stellar disk of Malin~2 extends far beyond that of J005042.73+002558.37, although both has almost the same stellar mass. Interestingly, J005042.73+002558.37 also has an $R_1$ radius of 58.8\,kpc, quite similar to Malin~2 (even slightly higher compared to the mean $R_1$ of 57.7\,kpc of Malin~2). This is the only galaxy in Fig. \ref{fig:mstar_r1} that has a higher $R_1$ than Malin~2. However, we can see that the radial extent of Malin~2 is well beyond the size of J005042.73+002558.37. This is expected as we see in the stellar mass surface density profile of Malin~2 from Fig. \ref{fig:malin2_color_density}, the stellar emission of Malin~2 extends well beyond and to almost twice its $R_1$. This also reinforces the idea that in both galaxies $R_1$ is a good proxy for the end of the in-situ star-forming disk, with one galaxy having no outer signatures of interactions (J005042.73+002558.37, J012015.34-002009.00) while the other (Malin~2) shows it.

We provide a more quantitative comparison in Fig. \ref{fig:malin2_comparison_full}, bottom panels. We show the stellar mass surface density profiles of Malin~2 along different wedges obtained in Sect. \ref{sect:color_stellar_mass_surface_density} in comparison to the profiles of spiral galaxies from \citet{Chamba2022}. The profiles are separated into different panels based on their stellar mass bins and types for a better illustration. Panel (a) of Fig. \ref{fig:malin2_comparison_full} shows the stellar mass surface density profiles of all the spiral galaxies from \citet{Chamba2022}, whereas panel (b) compares only Milky Way-mass spirals and panel (c) spirals of similar mass as Malin~2. Malin~2 stands out as an extreme case among all the galaxies shown and occupies the outermost envelope of the general profiles seen in other spiral galaxies (panel (a) of Fig. \ref{fig:malin2_comparison_full}). This is a different illustration of the extreme nature we see in the visual comparison of Malin~2 with other spirals in the top panels of Fig. \ref{fig:malin2_comparison_full}. Compared to both Milky Way-mass spirals and spirals with a similar mass to Malin~2, the stellar mass surface density profile of Malin~2 is extreme at all radii in its mass density and radial extent. In the case of panel (c) in Fig. \ref{fig:malin2_comparison_full}, the early drops seen in the 240$\degree$ and 330$\degree$ wedges of Malin~2 are closer to the general trend observed in the profiles of other galaxies in that mass range. Nevertheless, even in those wedge profiles, Malin~2 is towards the extreme side. The remaining wedge profiles (0$\degree$, 60$\degree$, 120$\degree$, and 180$\degree$) do not show any similarity with any of the profiles from other spiral galaxies. This indicates that the faint extended stellar emission seen in the outer disk of Malin~2 ($R>70$\,kpc) is likely to have originated from some past interactions rather than a secular origin.
However, it should be noted that the spiral sample does not have observations at the same depth as Malin~2  \citep[about 1 magnitude shallower than our observations;][]{Chamba2022}. Due to this difference in depth, we cannot rule out the presence of faint, extended features in the spiral sample of \citet{Chamba2022} at lower surface brightness levels as we reach in Malin~2. However, we note that the large radial extend of Malin 2 is not a result of the deeper imaging we obtain in this work. For instance, Fig. \ref{appendix:figure_decals_ttt_sigmastar_profile_comparison} shows a comparison of the stellar mass surface density profile of Malin~2 using DECaLS data and the profile we obatin in this work. We see that surveys like DECaLS, which is comparable to the depth of the data used in \citet{Chamba2022}, already show the large radial extend of Malin~2. However, the new faint features and asymmetric disk we see in this work are all almost invisible at the depth of those surveys.

\subsection{Comparison to other GLSB galaxies}\label{sect:comparison_to_other_glsbs}

\begin{figure*}[!htp]
  \centering
  \includegraphics[width=0.83\textwidth]{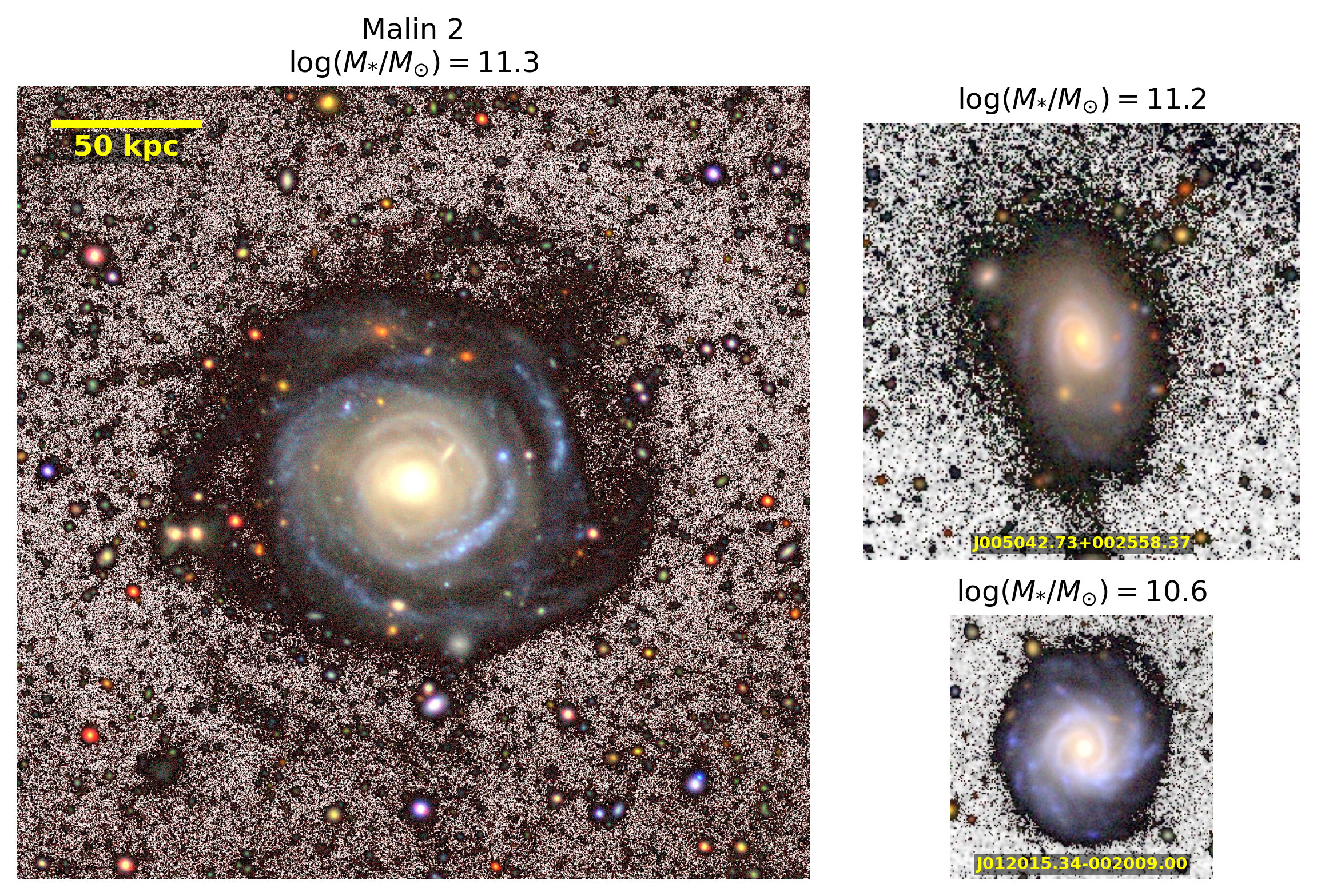}
  \\[0.5em]
  \includegraphics[width=0.83\textwidth]
  {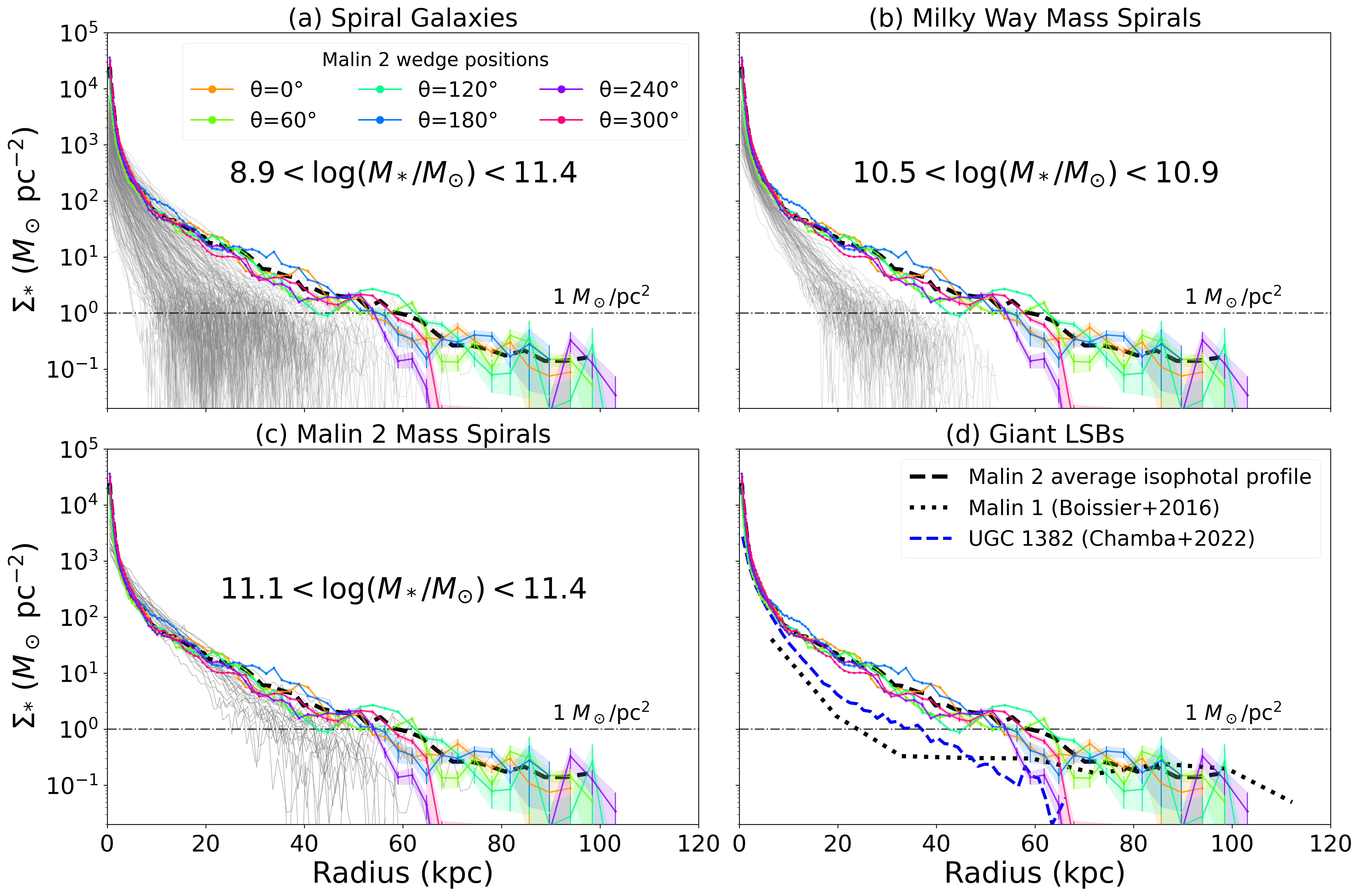}
  \caption{Comparison of Malin~2 images and stellar mass surface density profiles with other spiral galaxies. \textit{Top}: Visual size comparison showing Malin~2 (top left, log$(M_*/M_\odot) = 11.3$) compared to a spiral galaxy of a similar mass (J005042.73+002558.37; log$(M_*/M_\odot) = 11.2$ and a Milky Way-mass spiral (J012015.34-002009.00; $\log(M_*/M_\odot) = 10.6$). All the images have the same physical scale. The images of the galaxies in the right column are retrieved from the IAC Stripe82 Legacy Project \citep{Fliri2016,Roman2018}, which are about 1 magnitude shallower than our Malin~2 observations. \textit{Bottom}: Stellar mass surface density profiles comparing Malin~2's multi-directional wedge profiles (colored lines) with azimuthally averaged profiles of spiral galaxies from \citet{Chamba2022} (gray lines). The profiles are separated into four panels. Panel (a) is for all spirals, panel (b) Milky Way-mass spirals, and panel (c) Malin~2-mass spirals. All three panels are divided based on their stellar mass bins, as shown in the label. Panel (d) compares Malin~2 with other GLSB galaxies, namely Malin~1 \citep{Boissier2016} and UGC~1382 \citep{Chamba2022}. The dash-dotted horizontal lines mark the 1\,\msunpcsq{} level. The azimuthally averaged isophotal profile of Malin~2 is shown as black dashed line in all the panels.}
  \label{fig:malin2_comparison_full}
\end{figure*}

GLSB galaxies in the literature are generally characterized by their ``\textit{diffuseness index}'' ($d_i$) based on the \textit{B}-band disk scale length in kpc ($R_{\mathrm{s}}$) and central surface brightness in \magperarcsec{} ($\mu_{0,B}$), where $d_i = \mu_{0,B} + 5\log(R_{\mathrm{s}})$ \citep{Sprayberry1995,Hagen2016,Zhu2023,Bernaud2025}. A galaxy with $d_i > 27$ is defined as a giant LSB disk galaxy based on this diffuseness index criterion. \citet{Sprayberry1995} identify Malin~1, UGC~1382, UGC~6614, and Malin~2 as the most extreme cases of giant LSB disks with $d_i$ values of 35.2, 32.7, 29.7, and 29.4, respectively. Among them, Malin~2 has the lowest $d_i$, indicating it is the least extreme case in comparison to Malin~1 or UGC~1382 based on their diffuseness index. However, this picture changes if we take a deeper look into the panel (d) of Fig. \ref{fig:malin2_comparison_full}, where we compare the stellar mass surface densities of Malin~2 with those of Malin~1 from \citet{Boissier2016} and UGC~1382 from \citet{Chamba2022}. These two galaxies are the most extreme cases with the highest diffuseness index based on the \citet{Sprayberry1995} criterion. Based on panel (d) of Fig. \ref{fig:malin2_comparison_full}, Malin~1 and UGC~1382 have an $R_1$ radius of 26.5\,kpc and 34.7\,kpc, respectively. This is significantly smaller than the mean $R_1$ we obtain for Malin~2 (57.7\,kpc), making Malin~2 the most extended among them based on the $R_1$. Moreover, we should note that the diffuseness index criterion by \citet{Sprayberry1995} was formulated before the discovery of the extremely diffuse systems from current deep imaging surveys. Therefore, the diffuseness index criterion to select GLSB disk galaxies can also include some LSB dwarf galaxies like UDGs. For instance, the satellite candidate galaxy TTT-d1 we identified in Sect. \ref{sect:satellite} also satisfies the diffuseness index criterion if we assume a disk morphology and the $\mu_{0,g}$ and $R_{e}$ values from Table \ref{appendix_table:galfit_results}. This suggests that the definition of GLSB galaxies based on the diffuseness index may benefit from an update. Perhaps incorporating radii such as $R_1$ could lead to a more refined classification. However, this is beyond the scope of this work.

Malin~2 also shows interesting similarities (as well as some differences) to Malin~1 and UGC~1382 based on their stellar mass surface density profiles (see panel (d) of Fig. \ref{fig:malin2_comparison_full}). The inner part of their profiles ($R<10$\,kpc) appears to follow a similar trend\footnote{It should be noted that the Malin~1 stellar mass surface density profiles from \citet{Boissier2016}, shown in Fig. \ref{fig:malin2_comparison_full}, does not cover the very inner part of Malin~1 ($R < 6$\,kpc) due to the low-resolution multi-wavelength analysis performed in their work.}.
Between $10 < R < 60$\,kpc of their profiles, these three GLSB galaxies appear to be different, where Malin~2 has a large amount of stellar mass distributed within this radial range, and Malin~1 has the lowest. Close to the radius of $\sim$65\,kpc, the mass density profile of UGC~1382 appears to drop. This is quite similar to the drop in the 240$\degree$ and 300$\degree$ wedge profiles of Malin~2. Moreover, the UGC~1382 profile is well within the expected range found for other spiral galaxies, as we see in the previous panels of Fig. \ref{fig:malin2_comparison_full}. Beyond the radius of 70\,kpc (in panel \textit{d} of Fig. \ref{fig:malin2_comparison_full}), the stellar mass surface density profiles of both Malin~2 and Malin~1 look remarkably similar from 70 to 100\,kpc in radial range, in terms of their slope and extent. Both these profiles show a significant flattening in this range, with an almost constant stellar mass surface density of $\sim$0.3\,\msunpcsq{}. However, the flattening in the mass density profile of Malin~1 starts at a much smaller radius ($\sim$30\,kpc) compared to Malin~2 (which is only beyond 70\,kpc). \citet{Junais2024} found that the flattening of the extended disk in Malin~1 agrees with the radial gas metallically profile from VLT/MUSE observations, which is almost constant at about 0.6$Z_{\odot}$ in the extended disk of Malin~1, indicating the enrichment from a past/ongoing interaction. The similarity of the stellar mass density profiles of both Malin~2 and Malin~1 in their extended disks could point towards a similar scenario for Malin~2 as well. However, we require more observations to confirm this (e.g., kinematics and metallicity from IFU observations), which is beyond the scope of this work.

\subsection{Correlation of \Hi{} gas and stellar distribution: implications for the formation of Malin~2}\label{sect:hi_distribution}

\begin{figure}[!htb]
  \centering
  \includegraphics[width=0.48\textwidth]{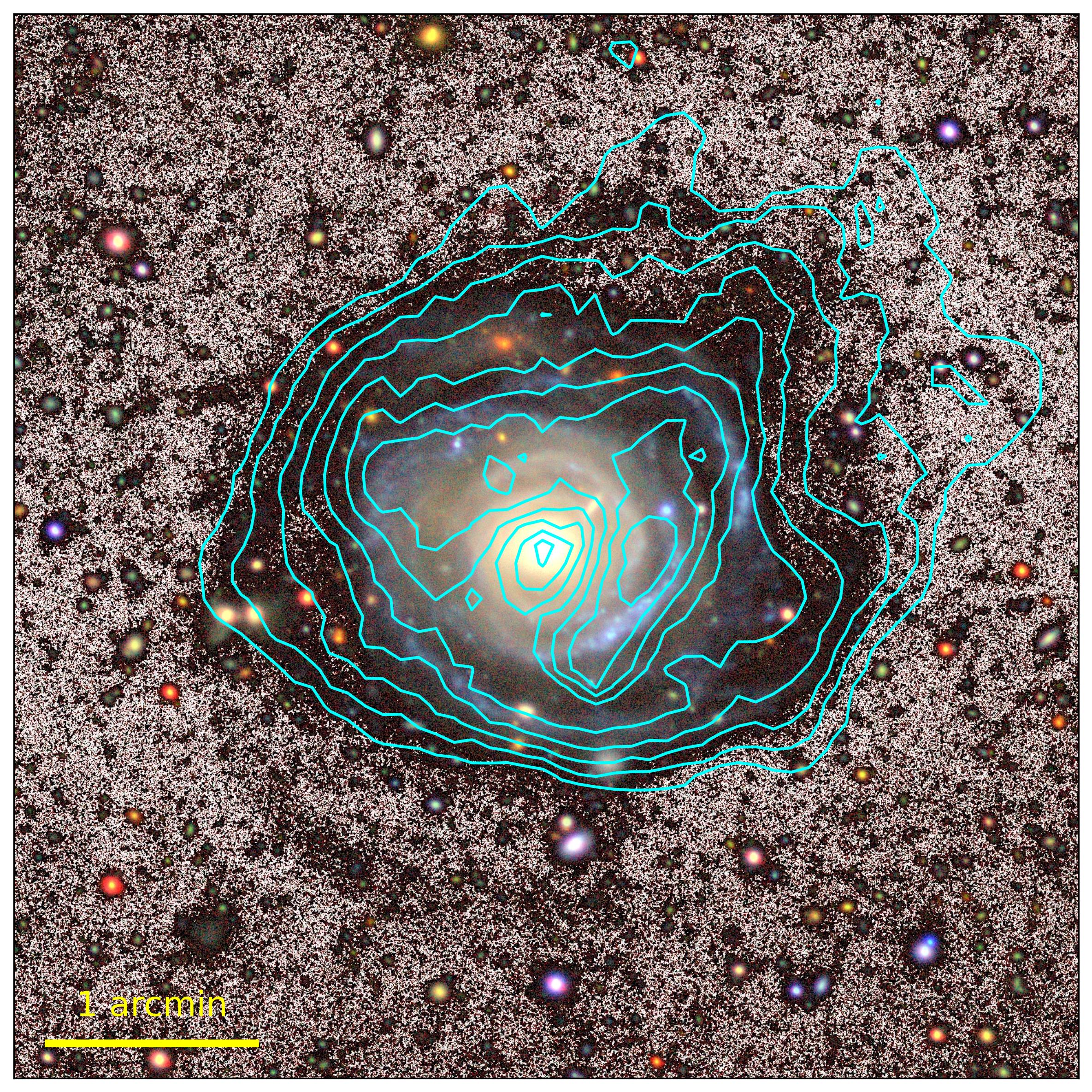}
  \caption{Deep optical image of Malin~2 overlaid with \Hi{} column density contours from \citet{Pickering1997} (cyan lines). The \Hi{} contours are spaced linearly with 10 contours ranging from a column density of $1.9\times10^{19}$\,cm$^{-2}$ to the peak column density of $1.9\times10^{20}$\,cm$^{-2}$.}
  \label{fig:malin2_hi_contours}
\end{figure}

We compare our deep imaging of Malin~2 with the \Hi{} gas distribution from \citet{Pickering1997} to investigate the relationship between the stellar and gas components in Malin~2's extended disk. Figure~\ref{fig:malin2_hi_contours} shows the overlay of \Hi{} contours on our deep optical image. The \Hi{} distribution shows a clear asymmetry, with a significant elongation towards the northwest compared to a more truncated southeastern extent. \citet{Pickering1997} noted that the northwest \Hi{} extension in Malin~2 is about 2 times longer than that in the southwest, creating a significantly lop-sided \Hi{} distribution (extending until 120\,kpc). Interestingly, the newly detected northwestern stellar overdensity in our imaging shows clear spatial coincidence with this extended \Hi{} emission. Moreover, the \Hi{} emission extends slightly beyond our observed diffuse stellar emission in this region. 

On the other hand, we do not see any \Hi{} emission associated with the stellar emission from the southeastern spiral-like structure or the dwarf galaxy (TTT-d1) visible in our imaging. Either these structures do not have any gas content within them, or they are too diffuse to be detected with the sensitivity of the VLA observations from \citet{Pickering1997}. In the case of TTT-d1, with its mean $g-r$ color of 0.53 mag (see Sect. \ref{sect:satellite}), it is likely that gas content in this dwarf galaxy (if exists) can only be detected at low \Hi{} column densities ($N_{\rm HI} \sim 10^{18}$\,cm$^{-2}$), as recently found by the \Hi{} detection of low surface brightness dwarf galaxies in the MONGHOOSE survey \citep{Maccagni2024}. Similarly, for the southeastern spiral-like structure, with its blue stellar population (see Table \ref{table:wedge_statistics} and Table \ref{appendix:table_malin2_regions_photometry}), the lack of detected \Hi{} emission might indicate a low column density gas ($N_{\rm HI} < 1.9 \times 10^{19}$\,cm$^{-2}$) below the sensitivity limits of the current VLA observations.

\citet{Jog2009} suggested that lopsided \Hi{} distribution in galaxies is likely a result of tidal encounters, gas accretion, or a global gravitational instability. The elongated \Hi{} morphology toward the northwest in Malin~2, together with the spatial overlap of the gas and stars in this region and the complex asymmetric stellar features revealed in our deep imaging, suggest that Malin~2's extreme size and structure is likely a result of contributions from tidal processes rather than a pure secular evolution. 
Although Malin~2 is located in a relatively low-density environment, which is generally the case with GLSB galaxies \citep{Saburova2021}, we identify two galaxies (LEDA~1638675 and LEDA~1635102) that are at a comoving distance of about 0.6 Mpc and 2.4 Mpc from Malin~2, respectively.
However, we do not see any additional evidence linking them in our optical imaging as well as the \Hi{} data. Additionally, there is also a galaxy group and cluster centered on the bright elliptical galaxies PGC~1645106 and PGC~032326, at a comoving distance of about 4 Mpc and 9 Mpc from Malin~2 towards the northeast direction, respectively. However, these distances are quite far to have any evident gravitational interactions. More follow-up observations are required to test if Malin~2 is experiencing any interactions or infalling towards these large scale structures.

%%%%%%%%%%%%%%%%%%%%%%%%%%%%%%%%%%%%%%%%%%%%%%%%%%%%%%%%%%%%%%
\section{Conclusions} \label{sect:conclusion}

We present deep, multi-band optical imaging of Malin~2, a prototype of GLSB galaxies. We use data from the recently commissioned Two-meter Twin Telescope (TTT) at the Teide Observatory to reveal the complex structure of Malin~2 at unprecedented surface brightness levels. Our main results are as follows:

\begin{itemize}
\item Our deep imaging reveals new stellar structures extending to $\sim110$\,kpc, including a northwestern overdensity that correlates spatially with \Hi{} emission and a southeastern spiral-like extension reaching $\mu_g \approx 30$\,\magperarcsec{}.

\item We report the discovery of a dwarf galaxy (TTT-d1) in the vicinity of Malin~2, at a projected distance of about 130\,kpc from its center. If confirmed to be at the distance of Malin~2, TTT-d1 will be the first satellite ultra-diffuse galaxy (UDG) around Malin~2.

\item A multi-directional wedge analysis of the surface photometry shows that Malin~2 has significant asymmetry in its stellar emission, color, and size along different directions. Such directional asymmetries would be missed by a traditional azimuthally averaged isophotal analysis.

\item By comparing Malin~2 to a large sample of nearby spiral galaxies and a few known GLSB galaxies, we find that Malin~2 lies at the extreme in its radial extent, larger and more extended, than any other nearby spiral or known GLSB galaxy.

\item The complex stellar features we identify in Malin~2, and their clear overlap with a lopsided \Hi{} gas distribution, suggest that Malin~2 experienced past interactions, which likely contributed to its GLSB disk formation.

\end{itemize}

Our findings highlight the importance of ultra-deep wide-field imaging for understanding the extreme population of GLSB galaxies. Recent deep imaging studies of other spiral galaxies have revealed complex substructure and satellite populations \citep[e.g.,][]{Taibi2025}. Our detection of a likely faint satellite around Malin~2 opens new avenues for constraining the formation histories of GLSB galaxies.  Future observations from surveys like LSST will allow the analysis of larger samples of GLSB galaxies to determine whether the complex structures we observe in Malin~2 are typical of this population.

%%%%%%%%%%%%%%%%%%%%%%%%%%%%%%%%%%%%%%%%%%%%%%%%%%%%%%%%%%%%%%

\begin{acknowledgements}
      We are grateful to Timothy Pickering who kindly shared us with the processed Malin~2 \Hi{} data cube. We would like to acknowledge interactions with Mpati Ramatsoku, Mathias Urbano, Gauri Sharma and Pablo Manuel Sánchez-Alarcón. This article is based on observations made in the Two-meter Twin Telescope (TTT\footnote{\url{http://ttt.iac.es}}) sited at the Teide Observatory of the Instituto de Astrofísica de Canarias (IAC), 
that Light Bridges operates in Tenerife, Canary Islands (Spain). The Observing Time Rights (DTO) 
used for this research were consumed in the PEI “GALAXDIF25". This research used storage and computing capacity in ASTRO POC's EDGE computing center at Tenerife 
under the form of Indefeasible Computer Rights (ICR). The ICR were consumed in the PEI "GALAXDIF25"
with the collaboration of Hewlett Packard Enterprise and VAST DATA. IT acknowledges support from the ACIISI, Consejer\'{i}a de Econom\'{i}a, Conocimiento y Empleo del Gobierno de Canarias and the European Regional Development Fund (ERDF) under a grant with reference PROID2021010044 and from the State Research Agency (AEI-MCINN) of the Spanish Ministry of Science and Innovation under the grant PID2022-140869NB-I00 and IAC project P/302302, financed by the Ministry of Science and Innovation, through the State Budget and by the Canary Islands Department of Economy, Knowledge, and Employment, through the Regional Budget of the Autonomous Community. This project is funded by the European Union (Widening Participation, ExGal-Twin, GA 101158446). J and JK acknowledge supplimentary funding from the European Union through the following grants: "UNDARK", GA 101159929 and MSCA EDUCADO, GA 101119830. Views and opinions expressed are however those of the author(s) only and do not necessarily reflect those of the European Union or European Research Executive Agency (REA). Neither the European Union nor the granting authority can be held responsible for them.
\end{acknowledgements}

\bibliographystyle{aa}
\bibliography{bibliography}

\begin{thebibliography}{53}
\expandafter\ifx\csname natexlab\endcsname\relax\def\natexlab#1{#1}\fi

\bibitem[{{Akhlaghi}(2019)}]{Akhlaghi2019}
{Akhlaghi}, M. 2019, in Astronomical Society of the Pacific Conference Series, Vol. 521, Astronomical Data Analysis Software and Systems XXVI, ed. M.~{Molinaro}, K.~{Shortridge}, \& F.~{Pasian}, 299

\bibitem[{{Akhlaghi} \& {Ichikawa}(2015)}]{Akhlaghi2015}
{Akhlaghi}, M. \& {Ichikawa}, T. 2015, \apjs, 220, 1

\bibitem[{{Alarcon} {et~al.}(2023){Alarcon}, {Licandro}, {Serra-Ricart}, {Joven}, {Gaitan}, \& {de Sousa}}]{Alarcon2023}
{Alarcon}, M.~R., {Licandro}, J., {Serra-Ricart}, M., {et~al.} 2023, \pasp, 135, 055001

\bibitem[{{Bakos} {et~al.}(2008){Bakos}, {Trujillo}, \& {Pohlen}}]{Bakos2008}
{Bakos}, J., {Trujillo}, I., \& {Pohlen}, M. 2008, \apjl, 683, L103

\bibitem[{{Barth}(2007)}]{Barth2007}
{Barth}, A.~J. 2007, \aj, 133, 1085

\bibitem[{{Bernaud} {et~al.}(2025){Bernaud}, {Boissier}, {Junais}, {Ma{\l}ek}, {Hugot}, \& {Galaz}}]{Bernaud2025}
{Bernaud}, E., {Boissier}, S., {Junais}, {et~al.} 2025, \aap, 700, A56

\bibitem[{{Bertin}(2006)}]{Bertin2006}
{Bertin}, E. 2006, in Astronomical Society of the Pacific Conference Series, Vol. 351, Astronomical Data Analysis Software and Systems XV, ed. C.~{Gabriel}, C.~{Arviset}, D.~{Ponz}, \& S.~{Enrique}, 112

\bibitem[{{Bertin} \& {Arnouts}(1996)}]{Bertin1996}
{Bertin}, E. \& {Arnouts}, S. 1996, \aaps, 117, 393

\bibitem[{{Bertin} {et~al.}(2002){Bertin}, {Mellier}, {Radovich}, {Missonnier}, {Didelon}, \& {Morin}}]{Bertin2002}
{Bertin}, E., {Mellier}, Y., {Radovich}, M., {et~al.} 2002, in Astronomical Society of the Pacific Conference Series, Vol. 281, Astronomical Data Analysis Software and Systems XI, ed. D.~A. {Bohlender}, D.~{Durand}, \& T.~H. {Handley}, 228

\bibitem[{{Boissier} {et~al.}(2016){Boissier}, {Boselli}, {Ferrarese}, {C{\^o}t{\'e}}, {Roehlly}, {Gwyn}, {Cuillandre}, {Roediger}, {Koda}, {Mu{\~n}os Mateos}, {Gil de Paz}, \& {Madore}}]{Boissier2016}
{Boissier}, S., {Boselli}, A., {Ferrarese}, L., {et~al.} 2016, \aap, 593, A126

\bibitem[{{Bothun} {et~al.}(1987){Bothun}, {Impey}, {Malin}, \& {Mould}}]{Bothun1987}
{Bothun}, G.~D., {Impey}, C.~D., {Malin}, D.~F., \& {Mould}, J.~R. 1987, \aj, 94, 23

\bibitem[{{Bradley} {et~al.}(2019){Bradley}, {Sipocz}, {Robitaille}, {Tollerud}, {Vin{\'\i}cius}, {Deil}, {Barbary}, {Busko}, {G{\"u}nther}, {Cara}, {Wilson}, {Conseil}, {Droettboom}, {Bostroem}, {Bray}, {Andersen Bratholm}, {Lim}, {Craig}, {Barentsen}, {Pascual}, {Donath}, {Greco}, {Perren}, {Kerzendorf}, {De Val-Borro}, {Dencheva}, {Albernaz De Ferreira}, {Souchereau}, {D'Eugenio}, \& {Weaver}}]{photutils}
{Bradley}, L., {Sipocz}, B., {Robitaille}, T., {et~al.} 2019, {astropy/photutils: v0.7.1}

\bibitem[{{Cardelli} {et~al.}(1989){Cardelli}, {Clayton}, \& {Mathis}}]{cardelli1989}
{Cardelli}, J.~A., {Clayton}, G.~C., \& {Mathis}, J.~S. 1989, \apj, 345, 245

\bibitem[{{Chamba} {et~al.}(2020){Chamba}, {Trujillo}, \& {Knapen}}]{Chamba2020}
{Chamba}, N., {Trujillo}, I., \& {Knapen}, J.~H. 2020, \aap, 633, L3

\bibitem[{{Chamba} {et~al.}(2022){Chamba}, {Trujillo}, \& {Knapen}}]{Chamba2022}
{Chamba}, N., {Trujillo}, I., \& {Knapen}, J.~H. 2022, \aap, 667, A87

\bibitem[{{Chambers} {et~al.}(2016){Chambers}, {Magnier}, {Metcalfe}, {Flewelling}, {Huber}, {Waters}, {Denneau}, {Draper}, {Farrow}, {Finkbeiner}, {Holmberg}, {Koppenhoefer}, {Price}, {Rest}, {Saglia}, {Schlafly}, {Smartt}, {Sweeney}, {Wainscoat}, {Burgett}, {Chastel}, {Grav}, {Heasley}, {Hodapp}, {Jedicke}, {Kaiser}, {Kudritzki}, {Luppino}, {Lupton}, {Monet}, {Morgan}, {Onaka}, {Shiao}, {Stubbs}, {Tonry}, {White}, {Ba{\~n}ados}, {Bell}, {Bender}, {Bernard}, {Boegner}, {Boffi}, {Botticella}, {Calamida}, {Casertano}, {Chen}, {Chen}, {Cole}, {Deacon}, {Frenk}, {Fitzsimmons}, {Gezari}, {Gibbs}, {Goessl}, {Goggia}, {Gourgue}, {Goldman}, {Grant}, {Grebel}, {Hambly}, {Hasinger}, {Heavens}, {Heckman}, {Henderson}, {Henning}, {Holman}, {Hopp}, {Ip}, {Isani}, {Jackson}, {Keyes}, {Koekemoer}, {Kotak}, {Le}, {Liska}, {Long}, {Lucey}, {Liu}, {Martin}, {Masci}, {McLean}, {Mindel}, {Misra}, {Morganson}, {Murphy}, {Obaika}, {Narayan}, {Nieto-Santisteban}, {Norberg}, {Peacock}, {Pier}, {Postman}, {Primak}, {Rae}, {Rai},
  {Riess}, {Riffeser}, {Rix}, {R{\"o}ser}, {Russel}, {Rutz}, {Schilbach}, {Schultz}, {Scolnic}, {Strolger}, {Szalay}, {Seitz}, {Small}, {Smith}, {Soderblom}, {Taylor}, {Thomson}, {Taylor}, {Thakar}, {Thiel}, {Thilker}, {Unger}, {Urata}, {Valenti}, {Wagner}, {Walder}, {Walter}, {Watters}, {Werner}, {Wood-Vasey}, \& {Wyse}}]{panstarrs}
{Chambers}, K.~C., {Magnier}, E.~A., {Metcalfe}, N., {et~al.} 2016, arXiv e-prints, arXiv:1612.05560

\bibitem[{{Das} {et~al.}(2010){Das}, {Boone}, \& {Viallefond}}]{Das2010}
{Das}, M., {Boone}, F., \& {Viallefond}, F. 2010, \aap, 523, A63

\bibitem[{{Fliri} \& {Trujillo}(2016)}]{Fliri2016}
{Fliri}, J. \& {Trujillo}, I. 2016, \mnras, 456, 1359

\bibitem[{{Gaia Collaboration} {et~al.}(2021){Gaia Collaboration}, {Brown}, {Vallenari}, {Prusti}, {de Bruijne}, {Babusiaux}, {Biermann}, {Creevey}, {Evans}, {Eyer}, {Hutton}, {Jansen}, {Jordi}, {Klioner}, {Lammers}, {Lindegren}, {Luri}, {Mignard}, {Panem}, {Pourbaix}, {Randich}, {Sartoretti}, {Soubiran}, {Walton}, {Arenou}, {Bailer-Jones}, {Bastian}, {Cropper}, {Drimmel}, {Katz}, {Lattanzi}, {van Leeuwen}, {Bakker}, {Cacciari}, {Casta{\~n}eda}, {De Angeli}, {Ducourant}, {Fabricius}, {Fouesneau}, {Fr{\'e}mat}, {Guerra}, {Guerrier}, {Guiraud}, {Jean-Antoine Piccolo}, {Masana}, {Messineo}, {Mowlavi}, {Nicolas}, {Nienartowicz}, {Pailler}, {Panuzzo}, {Riclet}, {Roux}, {Seabroke}, {Sordo}, {Tanga}, {Th{\'e}venin}, {Gracia-Abril}, {Portell}, {Teyssier}, {Altmann}, {Andrae}, {Bellas-Velidis}, {Benson}, {Berthier}, {Blomme}, {Brugaletta}, {Burgess}, {Busso}, {Carry}, {Cellino}, {Cheek}, {Clementini}, {Damerdji}, {Davidson}, {Delchambre}, {Dell'Oro}, {Fern{\'a}ndez-Hern{\'a}ndez}, {Galluccio}, {Garc{\'\i}a-Lario},
  {Garcia-Reinaldos}, {Gonz{\'a}lez-N{\'u}{\~n}ez}, {Gosset}, {Haigron}, {Halbwachs}, {Hambly}, {Harrison}, {Hatzidimitriou}, {Heiter}, {Hern{\'a}ndez}, {Hestroffer}, {Hodgkin}, {Holl}, {Jan{\ss}en}, {Jevardat de Fombelle}, {Jordan}, {Krone-Martins}, {Lanzafame}, {L{\"o}ffler}, {Lorca}, {Manteiga}, {Marchal}, {Marrese}, {Moitinho}, {Mora}, {Muinonen}, {Osborne}, {Pancino}, {Pauwels}, {Petit}, {Recio-Blanco}, {Richards}, {Riello}, {Rimoldini}, {Robin}, {Roegiers}, {Rybizki}, {Sarro}, {Siopis}, {Smith}, {Sozzetti}, {Ulla}, {Utrilla}, {van Leeuwen}, {van Reeven}, {Abbas}, {Abreu Aramburu}, {Accart}, {Aerts}, {Aguado}, {Ajaj}, {Altavilla}, {{\'A}lvarez}, {{\'A}lvarez Cid-Fuentes}, {Alves}, {Anderson}, {Anglada Varela}, {Antoja}, {Audard}, {Baines}, {Baker}, {Balaguer-N{\'u}{\~n}ez}, {Balbinot}, {Balog}, {Barache}, {Barbato}, {Barros}, {Barstow}, {Bartolom{\'e}}, {Bassilana}, {Bauchet}, {Baudesson-Stella}, {Becciani}, {Bellazzini}, {Bernet}, {Bertone}, {Bianchi}, {Blanco-Cuaresma}, {Boch}, {Bombrun}, {Bossini},
  {Bouquillon}, {Bragaglia}, {Bramante}, {Breedt}, {Bressan}, {Brouillet}, {Bucciarelli}, {Burlacu}, {Busonero}, {Butkevich}, {Buzzi}, {Caffau}, {Cancelliere}, {C{\'a}novas}, {Cantat-Gaudin}, {Carballo}, {Carlucci}, {Carnerero}, {Carrasco}, {Casamiquela}, {Castellani}, {Castro-Ginard}, {Castro Sampol}, {Chaoul}, {Charlot}, {Chemin}, {Chiavassa}, {Cioni}, {Comoretto}, {Cooper}, {Cornez}, {Cowell}, {Crifo}, {Crosta}, {Crowley}, {Dafonte}, {Dapergolas}, {David}, \& {David}}]{Gaia2021}
{Gaia Collaboration}, {Brown}, A.~G.~A., {Vallenari}, A., {et~al.} 2021, \aap, 649, A1

\bibitem[{{Galaz} {et~al.}(2015){Galaz}, {Milovic}, {Suc}, {Busta}, {Lizana}, {Infante}, \& {Royo}}]{Galaz2015}
{Galaz}, G., {Milovic}, C., {Suc}, V., {et~al.} 2015, \apjl, 815, L29

\bibitem[{{Gil de Paz} \& {Madore}(2005)}]{GildePaz2005}
{Gil de Paz}, A. \& {Madore}, B.~F. 2005, \apjs, 156, 345

\bibitem[{{Hagen} {et~al.}(2016){Hagen}, {Seibert}, {Hagen}, {Nyland}, {Neill}, {Treyer}, {Young}, {Rich}, \& {Madore}}]{Hagen2016}
{Hagen}, L. M.~Z., {Seibert}, M., {Hagen}, A., {et~al.} 2016, \apj, 826, 210

\bibitem[{{Infante-Sainz} \& {Akhlaghi}(2024)}]{Infante-Sainz2024}
{Infante-Sainz}, R. \& {Akhlaghi}, M. 2024, Research Notes of the American Astronomical Society, 8, 10

\bibitem[{{Jog} \& {Combes}(2009)}]{Jog2009}
{Jog}, C.~J. \& {Combes}, F. 2009, \physrep, 471, 75

\bibitem[{{Junais} {et~al.}(2022){Junais}, {Boissier}, {Boselli}, {Ferrarese}, {C{\^o}t{\'e}}, {Gwyn}, {Roediger}, {Lim}, {Peng}, {Cuillandre}, {Longobardi}, {Fossati}, {Hensler}, {Koda}, {Bautista}, {Boquien}, {Ma{\l}ek}, {Amram}, \& {Roehlly}}]{Junais2022}
{Junais}, {Boissier}, S., {Boselli}, A., {et~al.} 2022, \aap, 667, A76

\bibitem[{{Junais} {et~al.}(2024){Junais}, {Weilbacher}, {Epinat}, {Boissier}, {Galaz}, {Johnston}, {Puzia}, {Amram}, \& {Ma{\l}ek}}]{Junais2024}
{Junais}, {Weilbacher}, P.~M., {Epinat}, B., {et~al.} 2024, \aap, 681, A100

\bibitem[{{Kasparova} {et~al.}(2014){Kasparova}, {Saburova}, {Katkov}, {Chilingarian}, \& {Bizyaev}}]{Kasparova2014}
{Kasparova}, A.~V., {Saburova}, A.~S., {Katkov}, I.~Y., {Chilingarian}, I.~V., \& {Bizyaev}, D.~V. 2014, \mnras, 437, 3072

\bibitem[{{Lang} {et~al.}(2010){Lang}, {Hogg}, {Mierle}, {Blanton}, \& {Roweis}}]{Lang2010}
{Lang}, D., {Hogg}, D.~W., {Mierle}, K., {Blanton}, M., \& {Roweis}, S. 2010, \aj, 139, 1782

\bibitem[{{Maccagni} {et~al.}(2024){Maccagni}, {de Blok}, {Mancera Pi{\~n}a}, {Ragusa}, {Iodice}, {Spavone}, {McGaugh}, {Oman}, {Oosterloo}, {Koribalski}, {Kim}, {Adams}, {Amram}, {Bosma}, {Bigiel}, {Brinks}, {Chemin}, {Combes}, {Gibson}, {Healy}, {Holwerda}, {J{\'o}zsa}, {Kamphuis}, {Kleiner}, {Kurapati}, {Marasco}, {Spekkens}, {Veronese}, {Walter}, {Zabel}, \& {Zijlstra}}]{Maccagni2024}
{Maccagni}, F.~M., {de Blok}, W.~J.~G., {Mancera Pi{\~n}a}, P.~E., {et~al.} 2024, \aap, 690, A69

\bibitem[{{Martin} {et~al.}(2019){Martin}, {Kaviraj}, {Laigle}, {Devriendt}, {Jackson}, {Peirani}, {Dubois}, {Pichon}, \& {Slyz}}]{Martin2019}
{Martin}, G., {Kaviraj}, S., {Laigle}, C., {et~al.} 2019, \mnras, 485, 796

\bibitem[{{Mart{\'\i}nez-Lombilla} {et~al.}(2019){Mart{\'\i}nez-Lombilla}, {Trujillo}, \& {Knapen}}]{Martinez-Lombilla2019}
{Mart{\'\i}nez-Lombilla}, C., {Trujillo}, I., \& {Knapen}, J.~H. 2019, \mnras, 483, 664

\bibitem[{{Matthews} {et~al.}(2001){Matthews}, {van Driel}, \& {Monnier-Ragaigne}}]{Matthews2001}
{Matthews}, L.~D., {van Driel}, W., \& {Monnier-Ragaigne}, D. 2001, \aap, 365, 1

\bibitem[{{Montes} {et~al.}(2024){Montes}, {Trujillo}, {Karunakaran}, {Infante-Sainz}, {Spekkens}, {Golini}, {Beasley}, {Cebri{\'a}n}, {Chamba}, {D'Onofrio}, {Kelvin}, \& {Rom{\'a}n}}]{Montes2024}
{Montes}, M., {Trujillo}, I., {Karunakaran}, A., {et~al.} 2024, \aap, 681, A15

\bibitem[{{Moore} \& {Parker}(2006)}]{moore2006}
{Moore}, L. \& {Parker}, Q.~A. 2006, \pasa, 23, 165

\bibitem[{{Peng} {et~al.}(2010){Peng}, {Ho}, {Impey}, \& {Rix}}]{Peng2010}
{Peng}, C.~Y., {Ho}, L.~C., {Impey}, C.~D., \& {Rix}, H.-W. 2010, \aj, 139, 2097

\bibitem[{{Pickering} {et~al.}(1997){Pickering}, {Impey}, {van Gorkom}, \& {Bothun}}]{Pickering1997}
{Pickering}, T.~E., {Impey}, C.~D., {van Gorkom}, J.~H., \& {Bothun}, G.~D. 1997, \aj, 114, 1858

\bibitem[{{Roediger} \& {Courteau}(2015)}]{Roediger2015}
{Roediger}, J.~C. \& {Courteau}, S. 2015, \mnras, 452, 3209

\bibitem[{{Rom{\'a}n} \& {Trujillo}(2018)}]{Roman2018}
{Rom{\'a}n}, J. \& {Trujillo}, I. 2018, Research Notes of the American Astronomical Society, 2, 144

\bibitem[{{Rom{\'a}n} {et~al.}(2020){Rom{\'a}n}, {Trujillo}, \& {Montes}}]{Roman2020}
{Rom{\'a}n}, J., {Trujillo}, I., \& {Montes}, M. 2020, \aap, 644, A42

\bibitem[{{Saburova} {et~al.}(2021){Saburova}, {Chilingarian}, {Kasparova}, {Sil'chenko}, {Grishin}, {Katkov}, \& {Uklein}}]{Saburova2021}
{Saburova}, A.~S., {Chilingarian}, I.~V., {Kasparova}, A.~V., {et~al.} 2021, \mnras, 503, 830

\bibitem[{{Saburova} {et~al.}(2023){Saburova}, {Chilingarian}, {Kulier}, {Galaz}, {Grishin}, {Kasparova}, {Toptun}, \& {Katkov}}]{Saburova2023}
{Saburova}, A.~S., {Chilingarian}, I.~V., {Kulier}, A., {et~al.} 2023, \mnras, 520, L85

\bibitem[{{Schaye}(2004)}]{Schaye2004}
{Schaye}, J. 2004, \apj, 609, 667

\bibitem[{{Schlafly} \& {Finkbeiner}(2011)}]{Schlafly2011}
{Schlafly}, E.~F. \& {Finkbeiner}, D.~P. 2011, \apj, 737, 103

\bibitem[{{Schlegel} {et~al.}(1998){Schlegel}, {Finkbeiner}, \& {Davis}}]{Schlegel1998}
{Schlegel}, D.~J., {Finkbeiner}, D.~P., \& {Davis}, M. 1998, \apj, 500, 525

\bibitem[{{Sprayberry} {et~al.}(1995){Sprayberry}, {Impey}, {Bothun}, \& {Irwin}}]{Sprayberry1995}
{Sprayberry}, D., {Impey}, C.~D., {Bothun}, G.~D., \& {Irwin}, M.~J. 1995, \aj, 109, 558

\bibitem[{{Stone} {et~al.}(2021){Stone}, {Arora}, {Courteau}, \& {Cuillandre}}]{Stone2021}
{Stone}, C.~J., {Arora}, N., {Courteau}, S., \& {Cuillandre}, J.-C. 2021, \mnras, 508, 1870

\bibitem[{{Taibi} {et~al.}(2025){Taibi}, {Pawlowski}, {M{\"u}ller}, {B{\'\i}lek}, {J{\'u}lio}, {Kanehisa}, {Jovanovi{\'c}}, {Lalovi{\'c}}, \& {Samurovi{\'c}}}]{Taibi2025}
{Taibi}, S., {Pawlowski}, M.~S., {M{\"u}ller}, O., {et~al.} 2025, \aap, 699, A285

\bibitem[{{Teeninga} {et~al.}(2016){Teeninga}, {Moschini}, {Trager}, \& {Wilkinson}}]{Teeninga2016}
{Teeninga}, P., {Moschini}, U., {Trager}, S.~C., \& {Wilkinson}, M.~H.~F. 2016, Math. Morphol. Theory Appl., 1, 25

\bibitem[{{Trujillo} {et~al.}(2020){Trujillo}, {Chamba}, \& {Knapen}}]{Trujillo2020}
{Trujillo}, I., {Chamba}, N., \& {Knapen}, J.~H. 2020, \mnras, 493, 87

\bibitem[{{Trujillo} {et~al.}(2021){Trujillo}, {D'Onofrio}, {Zaritsky}, {Madrigal-Aguado}, {Chamba}, {Golini}, {Akhlaghi}, {Sharbaf}, {Infante-Sainz}, {Rom{\'a}n}, {Morales-Socorro}, {Sand}, \& {Martin}}]{Trujillo2021}
{Trujillo}, I., {D'Onofrio}, M., {Zaritsky}, D., {et~al.} 2021, \aap, 654, A40

\bibitem[{{Trujillo} \& {Fliri}(2016)}]{Trujillo2016}
{Trujillo}, I. \& {Fliri}, J. 2016, \apj, 823, 123

\bibitem[{{Zaritsky} {et~al.}(2024){Zaritsky}, {Golini}, {Donnerstein}, {Trujillo}, {Akhlaghi}, {Chamba}, {D'Onofrio}, {Eskandarlou}, {Hosseini-ShahiSavandi}, {Infante-Sainz}, {Martin}, {Montes}, {Rom{\'a}n}, {Sedighi}, \& {Sharbaf}}]{Zaritsky2024}
{Zaritsky}, D., {Golini}, G., {Donnerstein}, R., {et~al.} 2024, \aj, 168, 69

\bibitem[{{Zhu} {et~al.}(2023){Zhu}, {P{\'e}rez-Monta{\~n}o}, {Rodriguez-Gomez}, {Cervantes Sodi}, {Zjupa}, {Marinacci}, {Vogelsberger}, \& {Hernquist}}]{Zhu2023}
{Zhu}, Q., {P{\'e}rez-Monta{\~n}o}, L.~E., {Rodriguez-Gomez}, V., {et~al.} 2023, \mnras, 523, 3991

\end{thebibliography}
%%%%%%%%%%%%%%%%%%%%%%%%%%%%%%%%%%%%%%%%%%%%%%%%%%%%%%%%%%%%%%

\begin{appendix} 
\onecolumn

\section{Comparison of archival data with our new imaging of Malin~2}

\begin{figure*}[!htb]
  \centering
  \includegraphics[width=0.33\textwidth]{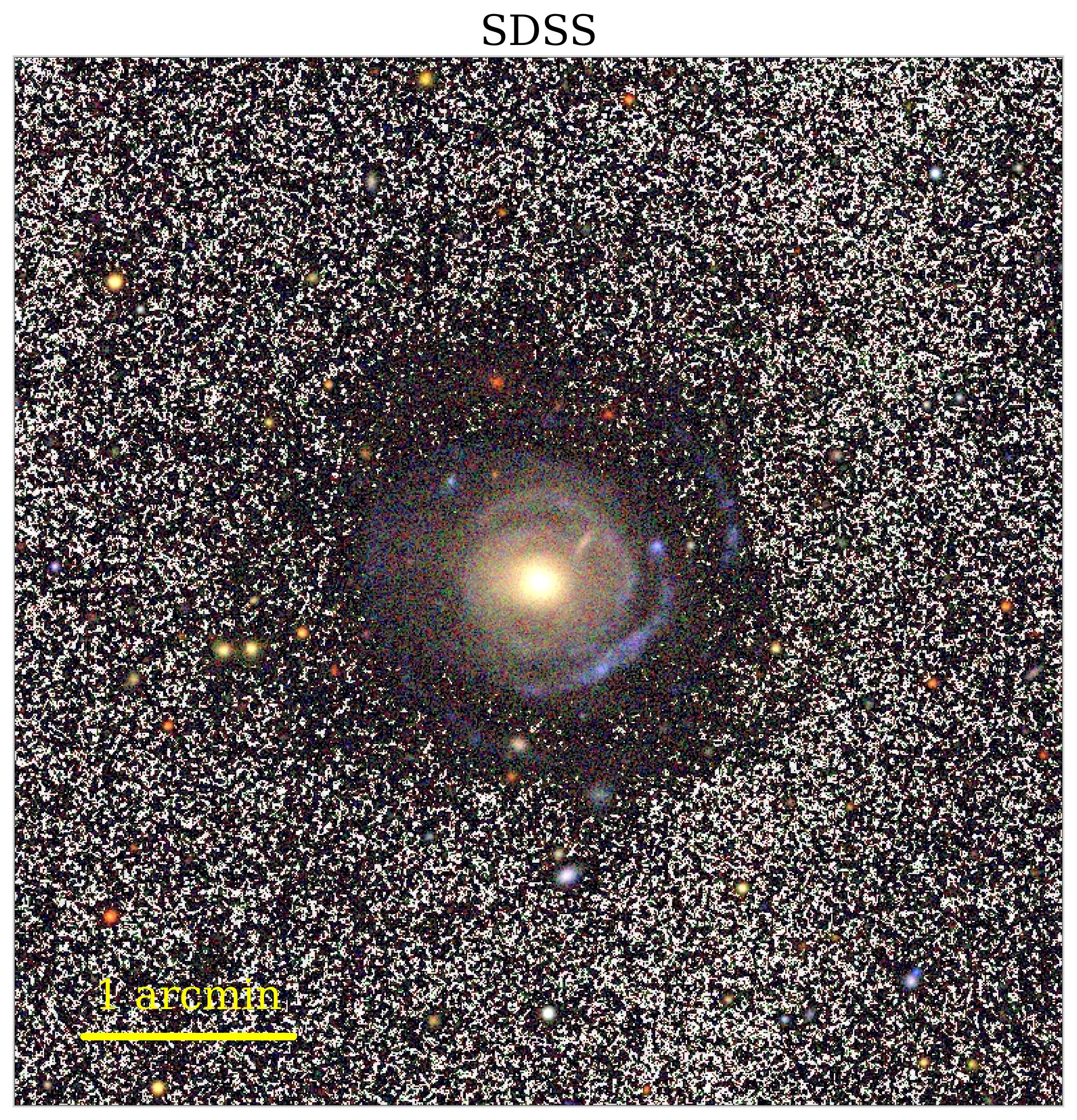}
  \includegraphics[width=0.33\textwidth]{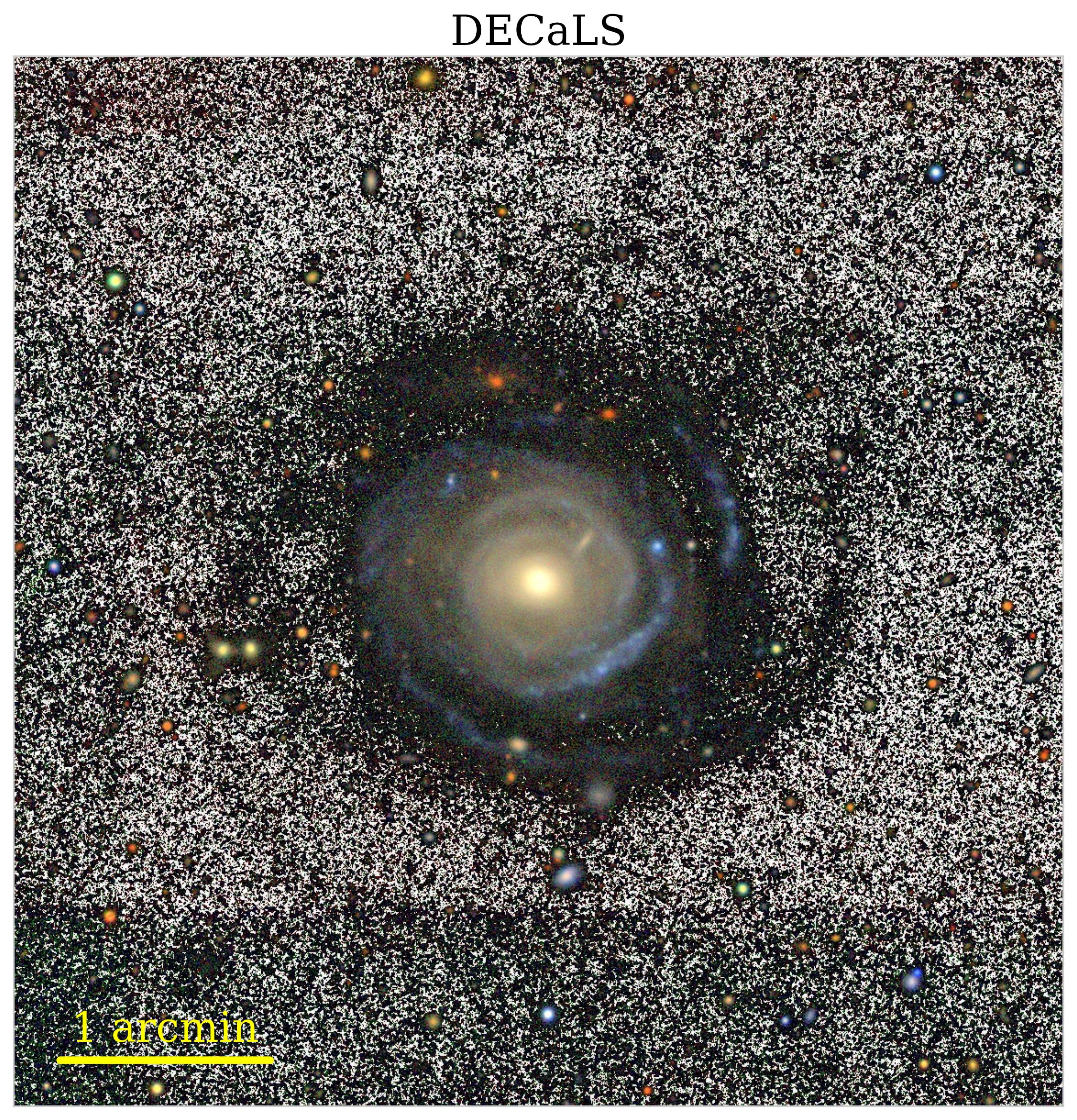}
  \includegraphics[width=0.33\textwidth]{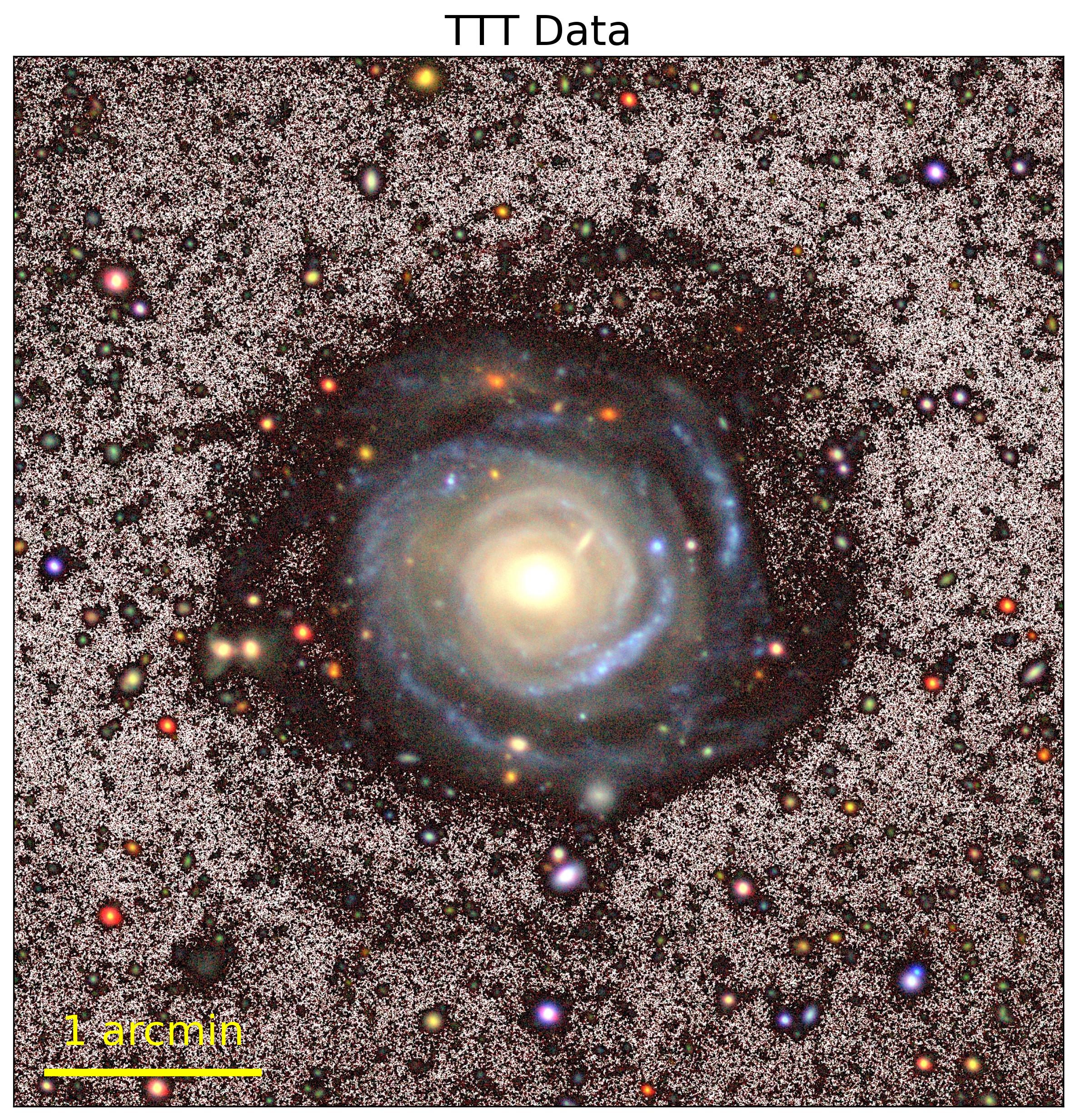}
  \caption{Comparison of Malin~2 images obtained from SDSS (left), DECaLS DR10 (middle) and the Two-meter Twin Telescope (TTT) at Teide Observatory (right). The color composite image is created using \textit{g}, \textit{r} and \textit{i}-band filters  with the \texttt{Gnuastro} package \texttt{astscript-color-faint-gray} \citep{Infante-Sainz2024}. The black and white background corresponds to the \textit{g}-band image. All the images are created using the same scale of contrast and zero point for a fair comparison. The images cover a field of view of $5\arcmin{} \times 5\arcmin{}$.}
  \label{appendix:figure_sdss_decals_ttt}
\end{figure*}

%************************************

\section{Aperture photometry of the diffuse regions in the extended disk} \label{appendix:aperture_photometry_regions}

\begin{figure}[!htb]
  \centering
  \includegraphics[width=\textwidth]{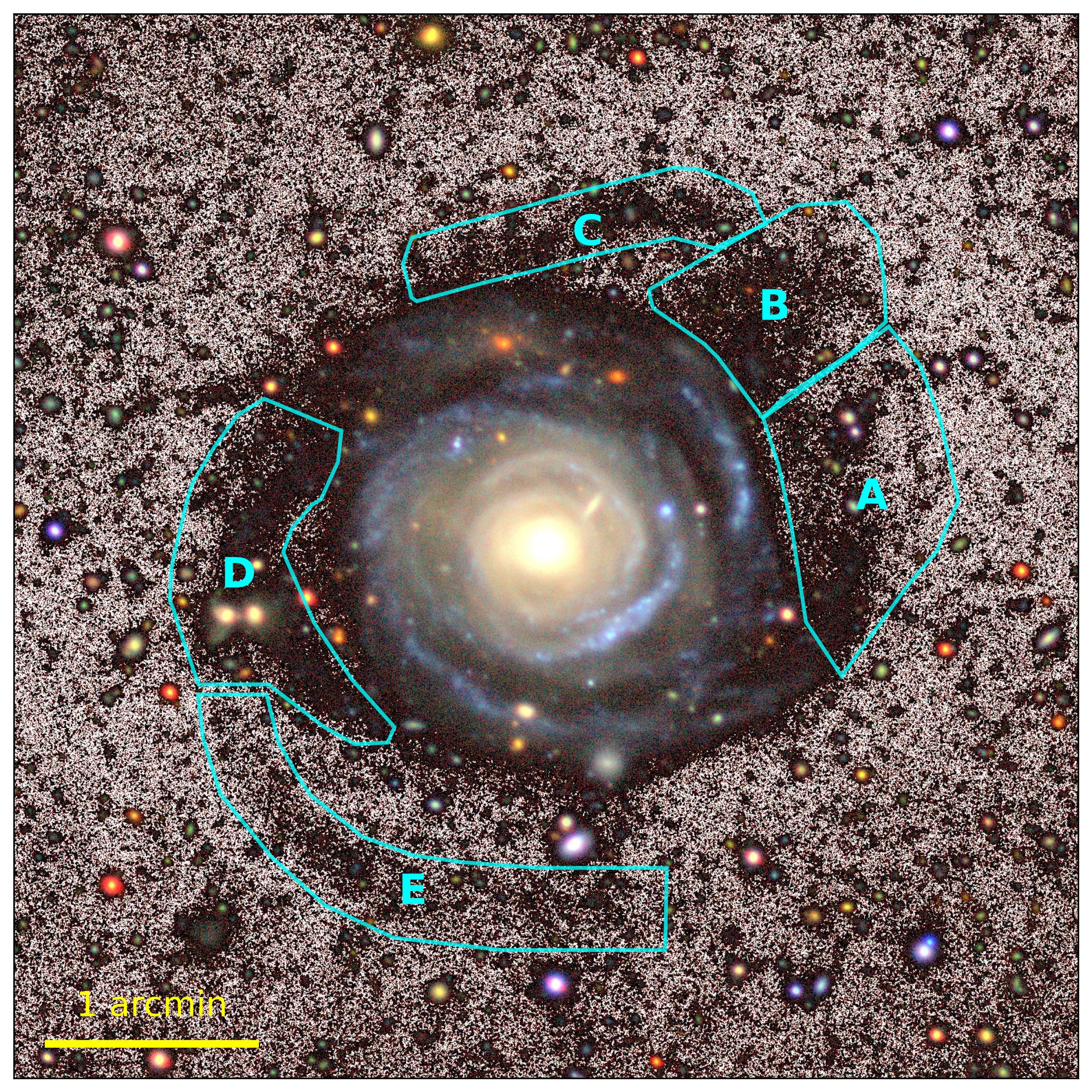}
  \caption{Color composite image of Malin~2 from TTT overlaid with visually identified regions of diffuse emission in the extended disk (cyan regions). See Table. \ref{appendix:table_malin2_regions_photometry} for the photometric properties of each labeled region.}
  \label{appendix:figure_regions_photometry}
\end{figure}

We perform aperture photometry measurements on several diffuse regions on the outermost part of the Malin~2 disk. Figure \ref{appendix:figure_regions_photometry} shows the regions that were visually selected based on the color image. We perform aperture photometry on these regions using the {\tt Photutils} \citep{photutils} package and the same mask as used in Sect. \ref{sect:sb_profile_measurements}. The background sky level and standard deviation for the measurements were estimated by placing a circular annulus of inner radius 130\arcsec{} and width of 30\arcsec{} around Malin~2, ensuring that it is outside the stellar emission from the galaxy. The results of the aperture photometry on these regions are given in Table \ref{appendix:table_malin2_regions_photometry}. Regions B and C have a similar $g-r$ color with a mean value of 0.45 mag, corresponding to the northwest diffuse region from Sect. \ref{sect:discovery_of_lsb_features} and the 60$\degree$ wedge from Sect. \ref{sect:sb_profile_measurements}. Regions D and E have a bluer color than regions B and C, with a mean $g-r$ of 0.23 mag. This indicates that the southeast spiral-arm-like structure seen in the optical image (corresponding to region E) has a young stellar population and is star-forming. Among all the 5 regions, Region A is the bluest. This is likely because a significant part of this region is an extension of the blue spiral arm visible in the optical image. 

\begin{table*}[!ht]
\centering
\caption{Aperture photometry results for diffuse substructures around Malin~2 as shown in Fig. \ref{appendix:figure_regions_photometry}.}
\begin{tabular}{lcccccc}
\hline
\hline
Region & Area & $\mu_g$ & $\mu_r$ & $m_g$ & $m_r$ & $g - r$ \\
       & (arcsec$^2$) & (mag arcsec$^{-2}$) & (mag arcsec$^{-2}$) & (mag) & (mag) & (mag) \\
       (1) & (2) & (3) & (4) & (5) & (6) & (7)\\

\hline
A & 3089.5 & 28.45 $\pm$ 0.01 & 28.37 $\pm$ 0.03 & 19.72 $\pm$ 0.01 & 19.64 $\pm$ 0.03 & 0.08 $\pm$ 0.03 \\
B & 2336.6 & 28.19 $\pm$ 0.01 & 27.79 $\pm$ 0.02 & 19.76 $\pm$ 0.01 & 19.37 $\pm$ 0.02 & 0.39 $\pm$ 0.02 \\
C & 1790.0 & 28.39 $\pm$ 0.02 & 27.90 $\pm$ 0.02 & 20.25 $\pm$ 0.02 & 19.77 $\pm$ 0.02 & 0.49 $\pm$ 0.03 \\
D & 2754.8 & 27.69 $\pm$ 0.01 & 27.44 $\pm$ 0.01 & 19.09 $\pm$ 0.01 & 18.84 $\pm$ 0.01 & 0.25 $\pm$ 0.01 \\
E & 3363.0 & 29.27 $\pm$ 0.03 & 29.05 $\pm$ 0.05 & 20.45 $\pm$ 0.03 & 20.23 $\pm$ 0.05 & 0.22 $\pm$ 0.05 \\
\hline
\end{tabular}
\tablefoot{(1) Region name; (2) Area of the region in arcsec$^2$, after masking; (3–4) Surface brightness of the region in $g$ and $r$-bands; (5–6) Total magnitudes in $g$ and $r$- bands; (7) $g - r$ color. All the magnitudes shown here are corrected for Galactic extinction.}
\label{appendix:table_malin2_regions_photometry}
\end{table*}

\newpage

%************************************

\section{Two-dimensional $g-r$ color map of Malin~2} \label{appendix:2d_g-r_map}

\begin{figure}[!htb]
  \centering
  \includegraphics[width=\textwidth]{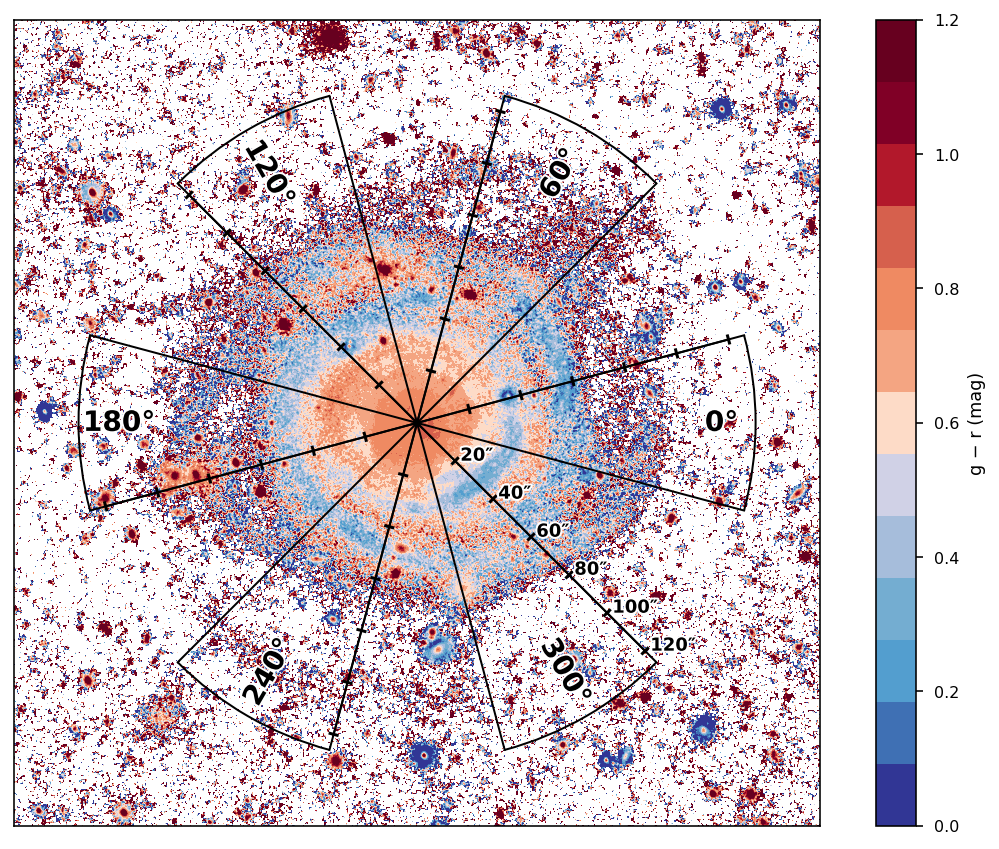}
  \caption{Two-dimensional $g-r$ color map of Malin 2 obtained from the TTT data. The color map is smoothed by a Gaussian with sigma of 1 pixel (0.194\arcsec{}) for a better visualisation of the LSB features. The black solid lines show the wedge positions used for the photometry in Sect. \ref{sect:sb_profile_measurements}. Six wedges, each with 30° opening angles are positioned at 0°, 60°, 120°, 180°, 240°, and 300°. A radial scalebar (in units of arcsec) is also shown along each wedge for easier comparison with the radial profiles from Sect. \ref{sect:sb_profile_measurements}.}
  \label{appendix:figure_2d_g-r_map}
\end{figure}

\newpage

\section{Photometry of the dwarf galaxy (TTT-d1) in the vicinity of Malin~2} \label{appendix:photometry_of_dwarf}

\begin{table*}[!htb]
\centering
\caption{Multi-band S\'ersic profile fitting results from {\tt GALFIT} \citep{Peng2010} for the dwarf galaxy (TTT-d1) near Malin~2.}
\label{tab:galfit-results}
\begin{tabular}{c c c c c c c c}
\hline
Filter & $R_{\mathrm{e}}$ & $n$ & $\mu_0$ & $\mu_{\mathrm{e}}$ & $m_{\text{tot}}$ & $b/a$ & PA \\
& (arcsec) & & (\magperarcsec{}) & (\magperarcsec{}) & (mag) & & (deg) \\
(1) & (2) & (3) & (4) & (5) & (6) & (7) & (8)\\
\hline
\textit{g} & 6.38 $\pm$ 0.22 & 0.86 $\pm$ 0.03 & 25.94 $\pm$ 0.09 & 27.45 $\pm$ 0.09 & 21.06 $\pm$ 0.04 & 0.79 & -46.74 \\
\textit{r} & 5.88 $\pm$ 0.23 & 0.86 $\pm$ 0.03 & 25.22 $\pm$ 0.10 & 26.74 $\pm$ 0.10 & 20.53 $\pm$ 0.04 & 0.78 & -35.95 \\
\hline
\end{tabular}
\tablefoot{(1) Name of the filter; (2) Effective radius; (3) S\'ersic index; (4) Central surface brightness; (5) Surface brightness at the effective radius; (6) Total apparent magnitude from the S\'ersic fit; (7) Axis ratio; (8) Position angle.}
\vspace{0.1cm} \label{appendix_table:galfit_results}
\end{table*}

\begin{figure}[!htb]
  \centering
  \includegraphics[width=0.48\textwidth]{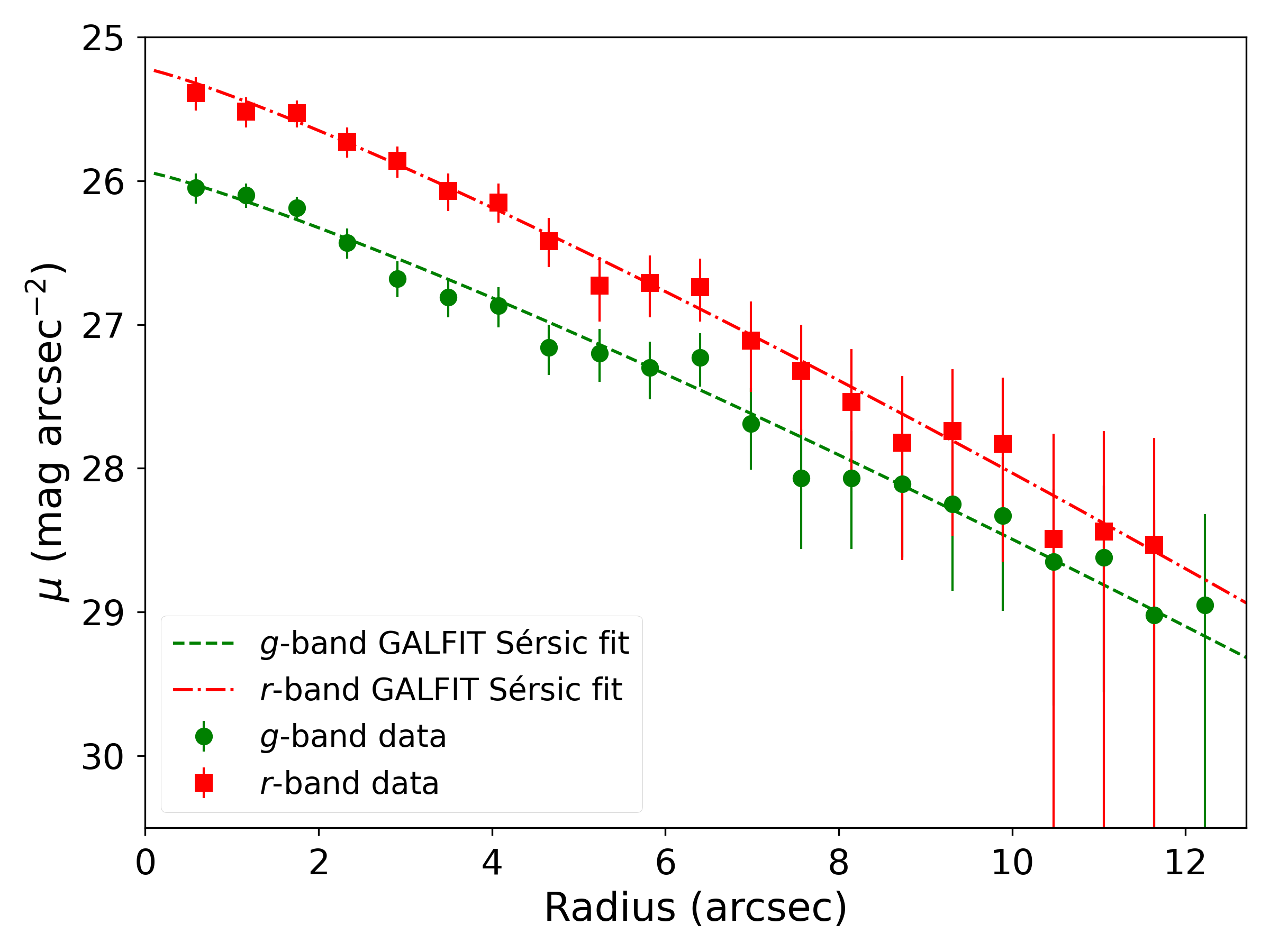}
  \caption{Radial surface brightness profiles of the dwarf galaxy (TTT-d1) near Malin~2. The \textit{g}- and \textit{r}-bands profiles are given in green circles and red squares, respectively. We measure the surface brightness profiles on elliptical isophotes using the {\tt Photutils} python package \citep{photutils} and following the method from \citet{Junais2022}. We fix the axis ratio and position angle for the profile measurements to the \textit{g}-band values from Table. \ref{appendix_table:galfit_results}. The green dashed line and the red dot-dashed lines are the best fit S\'ersic models (shown in Table. \ref{appendix_table:galfit_results}) independently obtained from the 2-dimensional fitting of the \textit{g}- and \textit{r}-band images using {\tt GALFIT} \citep{Peng2010}, respectively. The profiles shown here have not been corrected for inclination and foreground Galactic extinction.}
  \label{fig:dwarf_profiles} \label{appendix_fig:dwarf_profiles}
\end{figure}

\section{Stellar mass surface density profile comparison with TTT and DECaLS data}

\begin{figure*}[!htb]
  \centering
  \includegraphics[width=0.9\textwidth]{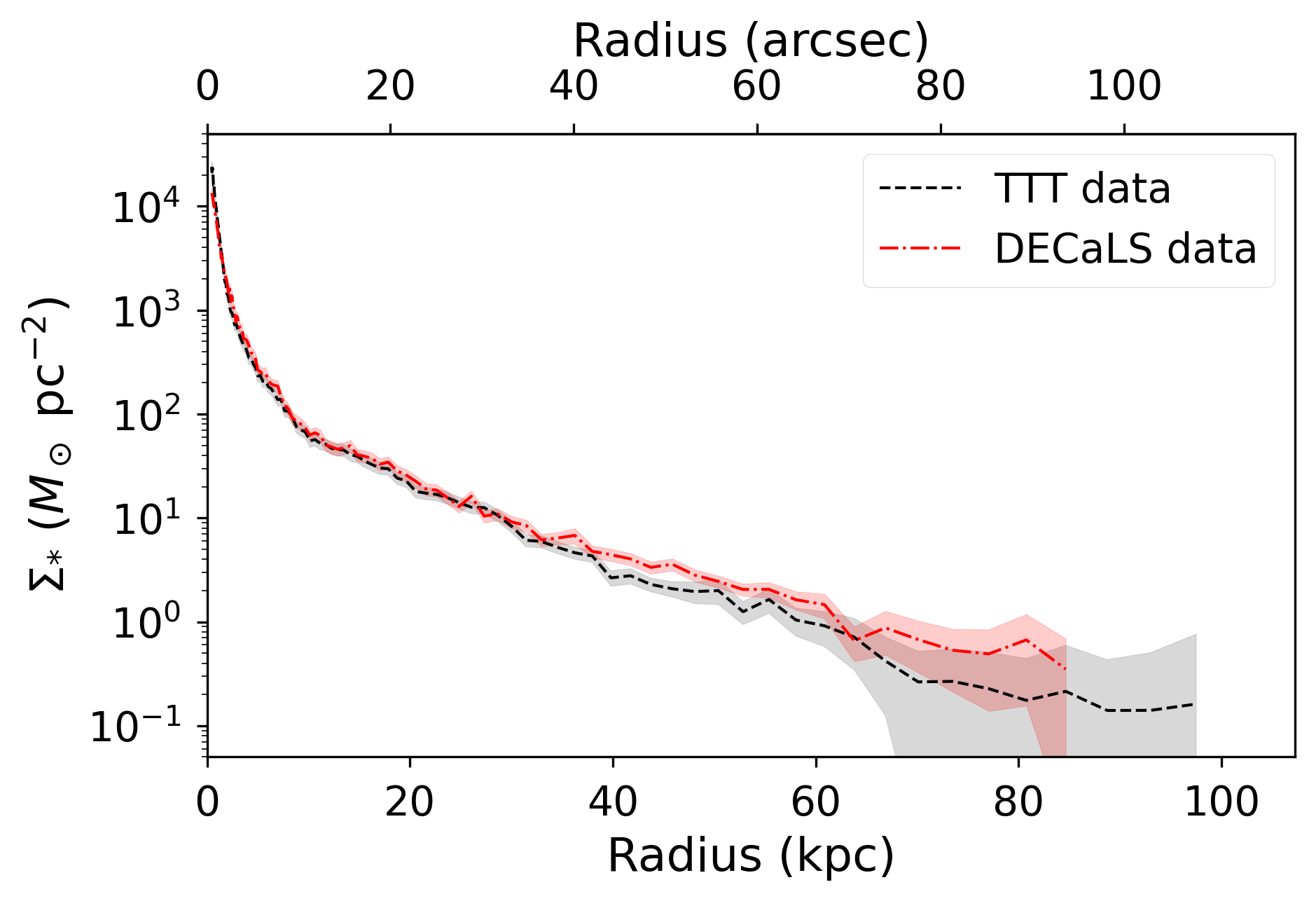}
  \caption{Comparison of the stellar mass surface density profiles of Malin~2 using the TTT data (black dashed line) and DECaLS data (red dash-dotted line). The shaded regions mark the $1\sigma$ uncertainty of the profiles. Both the profiles are measured following the same procedure as described in Sect. \ref{sect:color_stellar_mass_surface_density}. We see that both the profiles agree well, but shows some differences beyond 40 kpc radius which are still consistent within a $3\sigma$ level of uncertainty.}
  \label{appendix:figure_decals_ttt_sigmastar_profile_comparison}
\end{figure*}

\end{appendix}

\end{document}